\documentclass[fleqn,usenatbib]{mnras}

\usepackage[T1]{fontenc}

\DeclareRobustCommand{\VAN}[3]{#2}
\let\VANthebibliography\thebibliography
\def\thebibliography{\DeclareRobustCommand{\VAN}[3]{##3}\VANthebibliography}

\usepackage{graphicx}
\usepackage{amsxtra}
\usepackage{array,longtable}
\usepackage{journals}
\usepackage{url}
\usepackage{multirow}
\usepackage{latexsym}
\usepackage{amssymb}
\usepackage{hyperref}
\usepackage[singlelinecheck=false]{caption}
\bibpunct{(}{)}{;}{a}{}{,}
\usepackage{newtxtext,newtxmath}

\usepackage{ulem}

\usepackage{color}
\title
[Decomposition of S$^4$G galaxies with spiral arms]
{Galaxies decomposition with spiral arms -- II: 29 galaxies from S$^4$G}

\author[Chugunov et al.]
{Ilia V.~Chugunov$^1$\thanks{E-mail: chugunov21@list.ru},
Aleksandr V.~Mosenkov$^2$,
Alexander A.~Marchuk$^{1,3}$,
Sergey S.~Savchenko$^{1,3,4}$,
\newauthor
Ekaterina V.~Shishkina$^3$,
Maxim I.~Chazov$^4$,
Aleksandra E.~Nazarova$^4$,
Maria N.~Skryabina$^3$,
\newauthor
Polina I.~Smirnova$^3$,
Anton A.~Smirnov$^{1,3}$
\\
$^1$Central Astronomical Observatory of RAS, Russia\\
$^2$Department of Physics and Astronomy, N283 ESC, Brigham Young University, Provo, UT 84602, USA\\
$^3$St.Petersburg State University, 7/9 Universitetskaya nab., St.Petersburg, 199034, Russia\\
$^4$Special Astrophysical Observatory, Russian Academy of Sciences, 369167, Nizhnii Arkhyz, Russia
}

\date{Accepted XXX. Received YYY; in original form ZZZ}

\pubyear{2023}

\begin{document}
\label{firstpage}
\pagerange{\pageref{firstpage}--\pageref{lastpage}}

\maketitle
\begin{abstract}
Spiral structure can occupy a significant part of the galaxy, but properly accounting for it in photometric decomposition is rarely done. This may lead to significant errors in the parameters determined. To estimate how exactly neglecting the presence of spiral arms affects the estimation of galaxy decomposition parameters, we perform fitting of 29 galaxies considering spiral arms as a separate component. In this study, we utilize 3.6$\upmu$m-band images from the S$^4$G survey and use a new 2D photometric model where each spiral arm is modeled independently. In our model, the light distribution both along and across the arm can be varied significantly, as well as its overall shape. We analyze the differences between models with and without spiral arms, and show that neglecting spiral arms in decomposition causes errors in estimating the parameters of the disk, the bulge, and the bar. We retrieve different parameters of the spiral arms themselves, including their pitch angles, widths, and spiral-to-total luminosity ratio, and examine various relations between them and other galaxy parameters. In particular, we find that the spiral-to-total ratio is higher for galaxies with more luminous discs and with higher bulge-to-total ratios. We report that the pitch angle of spiral arms decreases with increasing bulge or bar fraction. We measure the width of the spiral arms to be 53\% of the disc scale length, on average. We examine the contribution of the spiral arms to the azimuthally-averaged brightness profile and find that spiral arms produce a ``bump'' on this profile with a typical height of 0.3--0.7 mag.

\end{abstract}


\section{Introduction}
The spiral arms of disc galaxies are remarkable structures with regions of ongoing star formation which are embedded into a fainter stellar disc. Studying spiral galaxies is of great importance as these galaxies represent a significant part (about 75\% of galaxies brighter than $M(B)=-20$~mag) of the local Universe (\citealt{Conselice2006}). 
Apart from spiral arms, the most prominent subsystems of disc galaxies, that are easily distinguishable in images, are a central spheroidal bulge and a flat extended stellar disc. A comprehensive analysis of the galactic structure requires measurements of the parameters (luminosities, spatial scales, etc.) of these subsystems. However, since physical components of a galaxy are embedded in each other and only a combined image of the sum of all components is observed, the structural analysis of the galaxies presents a complex computational problem. One possible solution to this issue is the so-called decomposition, which allows one to distinguish the light coming from the different galactic components.
\par
The key ingredient of the decomposition implies describing the surface brightness distribution of galactic components by analytical functions with specific parameters. The optimal values of these parameters can be found, and information about the physical components thus can be inferred (for example, see~\citealt{Erwin2015, Mendez-Abreu2017}). Due to its relative simplicity, bulge+disc decomposition is the most common approach to fit a galaxy image. For the bulge component, the de Vaucouleurs profile~\citep{deVaucouleurs1948} or the more general S{\'e}rsic function~\citep{Sersic1968} are often used, whereas the disc is usually described by an exponential profile suggested by~\citet{Freeman1970}. Bulge+disc decomposition has a moderate computational intensity and can be deployed as an automatic procedure. As a result, thousands and even millions of galaxies have been decomposed into several structural components~\citep{Simard2011, Bizyaev2014, Lang2016}. The obtained data allowed these authors to identify many physical scaling relations, such as the 
connection between the bulge fraction and the kinematics~\citep{Cappellari2013} or the supermassive black hole mass~\citep{Vika2012}, or relations between the bulge surface brightness, luminosity, and half-light radius~\citep{Fisher2010}, which appeared to be different for bulges and pseudobulges~\citep{Gadotti2009}.
\par
However, real galaxies often exhibit a more complex structure that cannot be reliably described by just two components, such as a disk and a bulge. Numerous studies have been focused on applying more advanced photometric models in galaxy decomposition. For example, the central regions often contain a bright component (a second bulge, a nuclear disc, etc.) and one should use an additional S{\'e}rsic profile~\citep{D'Souza2014, Erwin2021} to properly represent it in the model. Sometimes, the contribution of the active galactic nucleus is modeled using an unresolved point source~\citep{Gadotti2008} or a nuclear disc~\citep{Gadotti2020}. For disc components, more precise models are often adopted, with breaks of different types~\citep{Laine2014}, or with inner truncation justified by the presence of a bar or by quenching~\citep{Papaderos2022}. Additional thick or thin disc components and flaring can also be added~\citep{Mosenkov2021}. For edge-on galaxies, boxy or peanut-shaped (B/PS) bulges are often observed and are modelled via a separate component~\citep{Smirnov2020, Marchuk2022}. Despite the long list of aforementioned modifications, in some cases their inclusion may not be sufficient to accurately model a particular galaxy, since non-axisymmetric features may also be present~\citep{Peng2010}. 
In particular, spiral arms exhibit a wide variety of shapes with different winding tightness and width, thus making it difficult to properly account for spiral arms via photometric modelling. Moreover, the spiral structure is not always symmetric and the number of spiral arms varies from galaxy to galaxy. This diverse appearance is reflected in a number of classifications that were created to group them. \citet{Elmegreen1982} suggested a complex scheme in which there were a total of 12 different types. The classification was later revised in \citet{Elmegreen1987}, where the number of classes was reduced to 10 (for example, details of the bar presence were excluded from classification). In \citet{Elmegreen1990}, this classification was boiled down to three main classes that describe the general appearance of spirals. These include grand design galaxies that host two prominent spiral arms, multiple armed galaxies that have more than two distinguishable arms, and flocculent galaxies consisting of many fragmentary and fuzzy arms.
This complexity of spiral structure indicates that an analytical function for describing spiral arms in galaxies should be very flexible, have a rather large number of free parameters and, hence, be difficult to fit.
\par
Spirals are prominent, extended features in galaxy discs, so their contribution to the galaxy's luminosity is far from negligible, especially in grand design galaxies, where they account for up to 40\% of the total galactic luminosity~\citep{Savchenko2020}. Therefore, it is reasonable to expect that when we neglect their presence in our modeling, this may lead to systematic errors in the decomposition results. Moreover, due to the complexity of the appearance of spiral arms in galaxies, these errors are hard to estimate without a proper model for the spiral structure. The literature on this subject is scarce, and the existing studies are quite controversial. For example, some studies report that bulge parameters remain unchanged when spiral arms are added to model, while others state that the neglection of spiral arms causes significant biases in them~\citep{Lasker2014,Gao2017,Sonnenfeld2022,Lingard2020}.
\par
Some effort has been made to estimate inaccuracies of ``classical'' decomposition when spiral arms are not taken into account, by examining artificial images of galaxies~\citep{Lingard2020,Sonnenfeld2022}. Problems with this approach arise from the fact that mock galaxies look too simplified and unnatural when compared to real objects. Another approach used for decomposition is formulated in~\citet{Peng2010}. In the {\small{GALFIT}} framework, spiral arms are modelled as Fourier modes, modified by a rotation function of various forms. This approach is much more credible and robust even for spiral galaxies with complex structure. However, it is hard to use this model in practice, so this type of decomposition was applied only in a few studies with a handful of galaxies~\citep{Davis2012,Gao2017,Lasker2014}. Moreover, such a method is unable to properly trace pitch angle variations that are often observed in real galaxies~\citep{Savchenko2013,Savchenko2020}, and to produce a highly asymmetric spiral structure observed in many galaxies~\citep{Conselice1997}. In general, the results from the aforementioned studies, concerning the importance of spiral structure, indicate that the there is certainly room for improvement of the photometric model. 
\par
In this paper, we conduct a study of a sample of spiral galaxies using a new model of spiral arms, which is flexible enough to fit different forms of spiral structure. We aim to determine what kind of decomposition errors arise when spiral arms are not properly accounted for in decomposition. In addition, we are able to obtain the full parametric description of spiral arms, which is equally important. Since we use a model with the physically motivated parameters which are easy to interpret, this allows us to measure a plenty of parameters of spiral arms themselves, such as their widths or pitch angles. Measuring such parameters in the process of  decomposition has another advantage because the model of the spiral arms is fully treated in this case (see for comparison \citealt{Savchenko2020} who measured the parameters of spiral structure using residual images of spiral arms obtained after subtracting an azimuthally averaged model from the galaxy image).
\par
Our paper is organised as follows. In Section~\ref{sec:data}, we  describe the sample of galaxies and observational data used. In Section~\ref{sec:model}, we describe the model of spiral arm. Section~\ref{sec:decomposition} provides details about our decomposition and validation of our results. Our results concerning systematic errors of classical decomposition and statistics of the parameters of spiral arms (along with the general structural parameters) are discussed in Sections~\ref{sec:effect_on_other} and~\ref{sec:spiral_parameters}, respectively. We interpret our results in Section~\ref{sec:discussion} and summarize our findings and conclusions in Section~\ref{sec:conclusions}.

\section{The sample and data}
\label{sec:data}

For this study, we selected a sample of spiral galaxies from the Spitzer Survey of Stellar Structure in Galaxies (S$^4$G,~\citealt{Sheth2010}). S$^4$G contains more than 2300  nearby ($d < 40$~Mpc), bright ($M(B) < -15.5$~mag) and angularly extended ($D_{25} > 1$ arcmin) galaxies, located far from the Galactic plane ($|b| > 30^\circ$). S$^4$G provides images at 3.6~$\upmu$m and 4.5 $\upmu$m (we use only 3.6 $\upmu$m) with pixel size of $0.75$ arcsec and angular resolution about $2$ arcsec. In terms of depth, S$^4$G images reach $\mu_{3.6 \upmu \text{m}}$(AB)(1$\sigma$) $\sim 27$~mag/arcsec$^2$ (about 1~$M_\odot$/pc$^2$). S$^4$G is a suitable survey for this study because it covers a near-infrared part of the spectrum and has good resolution. Apart from that, spiral arms look relatively smooth in the infrared  \citep{1981ApJS...47..229E,1984ApJS...54..127E,1994A&A...288..365B}, drastically less affected by dust extinction, and therefore it is easier to fit them with an analytical model.

\par

Unfortunately, not all spiral galaxies in S$^4$G are suitable for the analysis with our decomposition model. The survey contains a large number of galaxies viewed at high inclination angles and, thus, their structure may be significantly distorted by the projection effects. Some spiral galaxies have too faint and/or flocculent  spiral structure to be reliably decomposed with our model of spiral arms. 
\par
To create a subsample of suitable spiral galaxies, we made two steps. At first, we considered a subsample of S$^4$G galaxies from~\citet{Diaz-Garcia2019}, which contains 391 not-highly inclined spiral galaxies, and selected only galaxies with low inclinations ($i > 40^\circ$). At the second stage, we performed visual inspection of all galaxies from the initial sample and selected objects with the most prominent spiral structure, which resulted in a sample of 29 galaxies. These galaxies are listed in Table~\ref{tab:sample_data} along with some of their basic properties.

\begin{table}
\caption{Some general characteristics of galaxies in our sample}
\begin{centering}
\label{tab:sample_data}
\begin{tabular}{lcccccc}
\hline\noalign{\smallskip}
Galaxy & $i$ & $d$ & $T$ & AC & $M(B)$ & $D_{25}$\\
& deg & Mpc & & & mag & arcmin \\
& (1) & (2) & (3) & (4) & (5) & (6) \\
\hline\noalign{\smallskip}
ESO508-024 & 31.5 & 42.9 & 5.0 & G & $-$19.3 & 2.1 \\
    IC0769 & 41.8 & 34.5 & 3.5 & G & $-$19.3 & 2.2 \\
    IC1993 & 21.3 & 13.6 & 2.0 & M & $-$18.2 & 2.8 \\
    IC2627 & 17.4 & 33.2 & 4.0 & G & $-$19.9 & 2.4 \\
    IC4237 & 46.1 & 40.1 & 3.0 & M & $-$19.8 & 2.1 \\
   NGC0895 & 44.3 & 27.8 & 5.0 & M & $-$19.9 & 3.3 \\
   NGC0986 & 26.0 & 24.8 & 2.0 & G & $-$20.4 & 4.0 \\
   NGC2460 & 45.7 & 20.7 & 1.0 & M & $-$19.0 & 1.8 \\
   NGC3507 & 29.8 & 17.9 & 3.0 & G & $-$19.2 & 3.0 \\
   NGC3596 & 21.4 & 20.9 & 4.0 & M & $-$19.8 & 3.6 \\
  NGC3683A & 47.8 & 35.3 & 4.0 & M & $-$20.0 & 2.0 \\
   NGC3684 & 47.3 & 20.4 & 5.0 & M & $-$19.2 & 2.3 \\
   NGC3686 & 35.2 & 20.3 & 4.0 & M & $-$19.5 & 2.9 \\
   NGC3687 & 18.0 & 37.9 & 1.5 & M & $-$19.9 & 1.4 \\
   NGC4067 & 41.9 & 37.4 & 2.0 & M & $-$19.5 & 1.1 \\
   NGC4165 & 48.3 & 29.9 & 2.5 & G & $-$18.1 & 1.2 \\
   NGC4314 & 20.4 & 17.0 & 1.0 & G & $-$19.8 & 3.7 \\
   NGC4548 & 39.0 & 11.0 & 1.5 & G & $-$19.3 & 5.5 \\
   NGC4680 & 39.3 & 38.3 & 3.0 & G & $-$19.7 & 1.3 \\
   NGC4902 & 21.6 & 40.6 & 2.5 & M & $-$21.2 & 2.7 \\
   NGC5194 & 32.9 & 8.6  & 4.0 & G & $-$21.1 & 13.8\\
   NGC5240 & 47.2 & 33.1 & 3.0 & M & $-$18.7 & 1.8 \\
   NGC5247 & 29.8 & 22.5 & 5.0 & G & $-$21.0 & 5.4 \\
   NGC5364 & 47.9 & 21.0 & 3.5 & M & $-$20.4 & 3.8 \\
   NGC5427 & 25.2 & 39.2 & 4.0 & G & $-$21.0 & 3.6 \\
   NGC7167 & 26.1 & 30.8 & 5.5 & G & $-$19.7 & 1.8 \\
   NGC7661 & 43.2 & 26.3 & 7.5 & G & $-$18.0 & 1.8 \\
   NGC7798 & 31.9 & 27.9 & 2.5 & G & $-$19.2 & 1.4 \\
 PGC028380 & 46.0 & 38.9 & 8.0 & G & $-$18.1 & 1.3 \\
\hline\noalign{\smallskip}
\end{tabular}\\
\end{centering}
(1) Galaxy inclination from~\citet{Salo2015};\\
(2) Distance from NED, \url{https://ned.ipac.caltech.edu/};\\
(3) Hubble type from~\citet{Buta2015};\\
(4) Arm class from~\citet{Buta2015}: G is grand-design, M is multi-armed;\\
(5) $B$-band absolute magnitude from~\citet{Makarov2014}, \url{http://leda.univ-lyon1.fr/};\\
(6) Optical diameter from~\citet{Makarov2014}\\
\end{table}

Besides galaxy images, S$^4$G provides error (noise) maps, a unified PSF image, and masks for each galaxy frame~\citep{Salo2015}. The PSF image has a FWHM of $1.7\arcsec$. Also, we note that mask images from S$^4$G only cover stellar objects and background/foreground galaxies. Different small-scale features in galaxies, such as star forming regions, were ignored. However these sources, if left unmasked, can influence fit results, and in some cases we had to modify the initial mask to exclude such objects from our fitting.

\section{Spiral arms model}
\label{sec:model}
We model each spiral arm independently with a function that has 21 free parameters. Almost the same model has been used in our other study devoted to the spiral structure of galaxy M\,51 (Marchuk et al., in prep.), where it was shown that our model reproduces well the observed properties of spirals (shape and light distribution). Here, we describe the basic equations and parameters of our model and refer the reader to Marchuk et al. (in prep.) for a detailed discussion of the reasons for the particular choice of functional form.
\par
The surface brightness distribution of an arm in polar coordinates $\left( r, \varphi\right)$ has a following form:
\begin{equation}
I(r, \varphi) = I_0 \times I_{\parallel}(r(\varphi), \varphi) \times I_{\bot}(r - r(\varphi), \varphi)\,,
\label{eq:model}
\end{equation}
where $I_0$ is the maximum brightness of the arm, and $r(\varphi)$ is the so-called shape function of the arm~(\citealt{Binney2008}, p. 471), which  determines the overall geometry of the arm. Two separate functions, $I_{\parallel}$ and $I_{\bot}$, determine how the surface brightness is distributed along and across the arm, respectively. All parameters of the model and their descriptions are listed in Table~\ref{tab:model_parameters}.

\par
The shape function $r(\varphi)$ is defined in such a way that $\log r$ is a fourth-degree polynomial in $\varphi$:
\begin{equation}
r(\varphi) = r_0 \times \exp \left(\varphi \sum_{i=0}^3 m_i\left(\frac{\varphi}{2 \pi}\right)^i\right)\,.
\end{equation}
 Here, $\varphi$ is counted from the angle $\varphi_0$ in the direction of spiral winding, ($r_0$, $\varphi_0$) is a starting point of the arm, and $m_i$ are the corresponding polynomial coefficients.
\par

\par
The distribution of the surface brightness along the spiral arm is a three-term function to allow the growth of the brightness at the beginning of the arm, an exponential decrease outside of its brightness point, and a truncation at the end of the arm:
\begin{equation}
I_{\parallel}(r(\varphi), \varphi) = \frac{1}{\bar{I}} \left(h_\text{s}\varrho(\varphi) \right)^{\varrho(\varphi_\text{max})} \times  e^{ -\varrho(\varphi) } \times T(\varphi_\text{cutoff}, \varphi_\text{end})\,,
\end{equation}
where ${\bar{I}}$ is a normalisation constant, $h_\text{s}$ is the scale of exponential decay, $\varrho=(r(\varphi)-r_0)/h_\text{s}$ is a relative distance to the origin of the arm, $\varphi_\text{max}$ is the value of the polar angle where the maximum surface brightness $I_0$ is reached. The truncation function
\begin{equation}
    T(\varphi_\text{cutoff}, \varphi_\text{end}) =  1-\Theta(\varphi-\varphi_\text{cutoff})\frac{\varphi-\varphi_\mathrm{cutoff}}{\varphi_\text{end}-\varphi_\text{cutoff}}
\end{equation}
depends on the value of the angle $\varphi_\text{cutoff}$, where the profile cutoff begins, and $\varphi_\text{end}$, where and after which the intensity is set to zero. $\Theta$ is the Heaviside function.
\par 
A surface brightness distribution across the arm is modelled by a couple of S{\'e}rsic functions to fit the inside (closer to the galaxy center) and the outside (farther from the galaxy center) parts of the arm. Both functions have an additional common parameter, that governs how their widths changes along the arm to allow a spiral pattern with variable width:
\begin{equation}
\label{eq:I_bot}
I_{\bot}^\text{in/out}(\rho,\varphi) = \exp \left(-b_\text{n}^\text{in/out} \times \left(\frac{\rho}{\sqrt{(w_\text{e}^\text{in/out})^2 + \left(\varphi \times \xi \right)^3}} \right)^{\frac{1}{n^\text{in/out}}}\right),
\end{equation}
where upper indexes ``in'' and ``out'' stand for the inner  and outer parts of the arm. Here, $\rho=r(\varphi) -r $ is the radial distance to the ridge of the arm, $n$ is the S{\'e}rsic index, $w_\text{e}$ is the half-width of the spiral, the coefficient $\xi$ describes the widening of the arm towards the outer edge of the galaxy, and $b_\text{n}$ is not a free parameter, but a normalization coefficient of the S{\'e}rsic law.

\par 
Some notable features of our model are
\begin{enumerate}
    \item each spiral arm can be fitted individually (no imposed symmetry around the centre, and an arbitrary number of spiral arms can be included in the model)
    \item The major advantage of the described model is that almost all parameters have a clear geometrical/physical meaning. Although this number of parameters seems to be large, it is necessary to make the model flexible enough to reproduce various shapes and properties of spiral arms. For example, the model of spiral arms in~\citet{Peng2010} contains up to 103 parameters which mostly have no physical meaning by themselves. On the contrary, in~\citet{Lingard2020}, a very simple model of 6 parameters was used. In particular, their model produces spiral arms with a constant pitch angle which is a crude approximation for real galaxies, especially in their periphery \citep{Savchenko2013,Savchenko2020}.
    
    \item the pitch angle $\upmu_\varphi$ at a certain point on the arm with azimuthal angle $\varphi$ can be calculated as follows:
\begin{equation}
\upmu_\varphi = \arctan \left( \sum_{i=0}^3 (i + 1) m_i \left(\frac{\varphi}{2 \pi} \right)^i\, \right).
\end{equation}
    \item The average pitch angle $\langle\upmu\rangle$ for any part of the arm (for example, in the range $[\varphi_1, \varphi_2]$) can be found as the arctangent of the slope coefficient of the linear fit of the points of the spiral structure in the log-polar coordinates, i.e. $\log r(\varphi_2)/r(\varphi_1) = \tan \langle\upmu\rangle \times (\varphi_2 - \varphi_1)$.
    
\end{enumerate}

\par



The centre point ($X_0, Y_0$), inclination $i$, and positional angle PA are also amidst the parameters of the model and correspond to the position and orientation of the galaxy as a whole. The centre point and inclination are usually the same for all components (disc, bulge, spirals). 

\begin{table}
\begin{centering}
\caption{Parameters of the spiral arm model.}
\label{tab:model_parameters}
\begin{tabular}{lcc}
    \hline\noalign{\smallskip}
	Part & Parameter & Description \\
    \hline\noalign{\smallskip}
	  & $X_0, Y_0$* & Coordinates of the galactic center \\
	& PA* & Position angle of galactic plane \\
	  & $i$* & Inclination of galactic plane \\
	  & $I_0$ & Maximum intensity in the arm \\
    \hline\noalign{\smallskip}
	\multirow{3}{*}{$r(\varphi)$} & $m_{0 \ldots 3}$* & Pitch angle polynomial coefficients \\
	  & $r_0, \varphi_0$* & Coordinates of beginning of the arm \\
	  & cw/ccw* & Arm winding direction \\
    \hline\noalign{\smallskip}
	\multirow{4}{*}{$I_\parallel$} & $\varphi_\text{max}$ & Azimuthal angle of maximum intensity \\
	  & $\varphi_\text{cutoff}$* & Azimuthal angle of cutoff beginning \\
	  & $\varphi_\text{end}$* & Azimuthal angle of end of the arm \\
	  & $h_\text{s}$ & Arm exponential scale \\
    \hline\noalign{\smallskip}
	\multirow{3}{*}{$I_\bot$} & $w_\text{e}^\text{in}, w_\text{e}^\text{out}$ & Half-width inwards and outwards \\
	  & $n^\text{in}, n^\text{out}$ & S{\'e}rsic index inwards and outwards \\
	  & $\xi$ & Arm width increase rate \\ 
    \hline\noalign{\smallskip}
\end{tabular}
\end{centering}\\
\textit{Notes}: parameters marked with the asterisk were determined at the preliminary step and were fixed during the decomposition (see text).
\end{table}

\section{Decomposititon and validation}
\label{sec:decomposition}

For decomposition of our galaxy images, we employ the {\small{IMFIT}} package~\citep{Erwin2015} which is flexible enough to add new user-defined classes and to use different optimization techniques. We modified the latest {\small{IMFIT}} version 1.9 to implement our spiral arm model function\footnote{This new class incorporated in {\small{IMFIT}} can be found at \url{https://github.com/IVChugunov/IMFIT_spirals}.}, described in detail Sec.~\ref{sec:model}.

\par

The main difficulty of using a complex model with a big number of degrees of freedom is finding proper initial values of the free parameters. If initial values are not close enough to their optimal values, the optimisation iteration process can converge to one of possible local minimums or even to singular points of the function used (one or more parameters can become equal to zero or infinity). Moreover, fitting of a large number of parameters is a time-consuming procedure and finding an appropriate initial guess becomes even more important. An additional problem is to find a proper combination of model components, which better suits a particular galaxy, because galaxies can have a different set of structures (for example, a bar, a lens, an active nucleus may or may not appear in a galaxy).

\par

At a first step, we perform a decomposition using a model without spiral arms, in order to use its results as initial conditions for a more complicated model with spiral arms. As a starting point for this decomposition, we use results from~\citet{Salo2015}, where such a decomposition was performed using the {\small{GALFIT}} code \citep{Peng2002}. For each galaxy in our sample, we converted their results into an {\small{IMFIT}} input file and then performed {\small{IMFIT}} fitting. In some cases, we had to adopt a different set of components, e.g. an exponential disc + point source model instead of an exponential disc + S\'{e}rsic bulge model (in case the bulge size is too small for a proper fitting), or adopt a more complex structure by adding new components.

\par

To simplify our fitting, we obtained some spiral arm parameters separately and fixed them, namely parameters which define position, shape and length of spiral arms. To determine these parameters, we traced the spiral arms by manually placing points in SAOImageDS9\footnote{\url{https://sites.google.com/cfa.harvard.edu/saoimageds9}} package~\citep{Joye2003} along the arms ridges and then saved their coordinates. Our Python script used these coordinates for estimating initial guess values and then performed approximation of the shape of the spiral arms by means of our model. The initial guess for the remaining parameters of the spiral arm model was chosen manually and fitted by Levenberg-Marquardt algorithm~\citep{More1978} which is one of the fitting methods implemented in {\small{IMFIT}}. Since the initial guess and bounds of parameters were not always good, we usually had to perform fitting multiple times, especially for the models with spiral arms. We estimated the goodness of fit by visual inspection of the real images and their models, as well as the corresponding residual images and azimuthally averaged surface brightness profile. Adding spirals to our fit model resulted in a considerably longer computational process, typically 1--2 orders of magnitude longer as compared to a simple fit with a standard set of bulge and disc components. On average, fitting of a single model without spiral arms required computational time of about tens of seconds with Intel Core i5-7200U CPU. When spiral arms were added to the model, the time were from a few minutes up to an hour. As mentioned above, to obtain a final model of a single galaxy it was usually needed to perform fitting several times, increasing the required time even more.

\par

\begin{figure*} 
   \centering
   \includegraphics[width=1.95\columnwidth, angle=0]{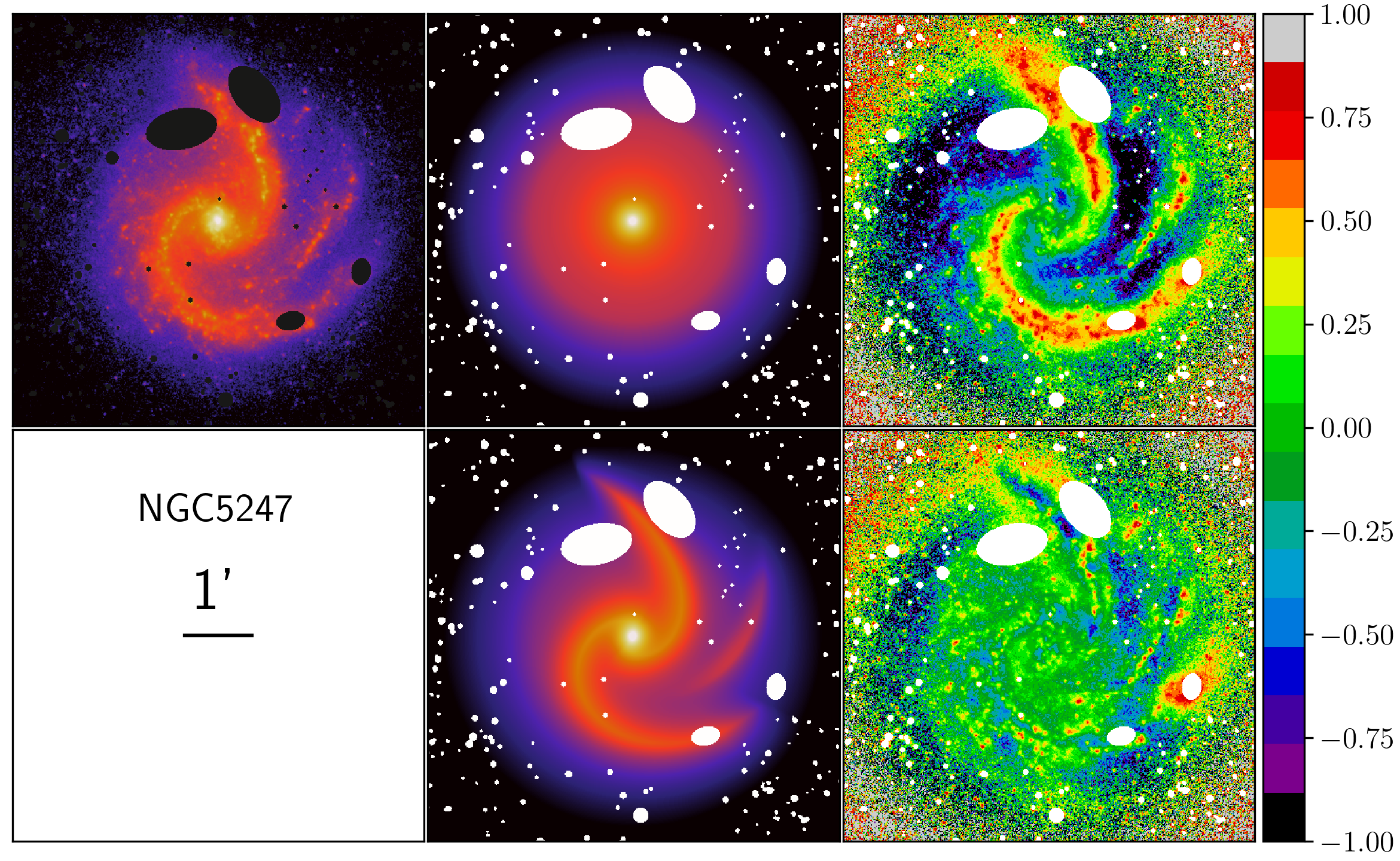}
   \caption{Decomposition of NGC~5247 shown as an example. From left to right: original image, models, and relative residuals. The classical model and thecorresponding residual image are on the top row, model with spiral arms and corresponding resudual are on the bottom row.\\
   Red and blue colors on the residual image depict under- and overestimated brightnesses in the model image, respectively, whereas the green color means good agreement.} 
   \label{fig:NGC5247}
\end{figure*}

At the end we obtained two models for each galaxy: the first one consists of only axially symmetric components without spiral arms, hereafter ``classical model'', and the other model contains, besides the same set of components, an appropriate number of well-visible spiral arms. This approach allows us to compare these models with each other and to estimate biases in the measured parameters if spirals are not included into the model. In Fig.~\ref{fig:NGC5247}, we show an example of decomposition for NGC~5427. Decomposition results for all galaxies, including model parameters and images, are avaliable online; example for one galaxy is shown in~\hyperref[sec:appendix]{Appendix A} (Fig.~\ref{fig:appendix_example}). Large inconsistencies between the classical model and the input galaxy image are clearly seen on the residual image, whereas the model with spiral arms has much more resemblance with the original galaxy image.

\begin{figure} 
   \centering
   \includegraphics[width=0.95\columnwidth, angle=0]{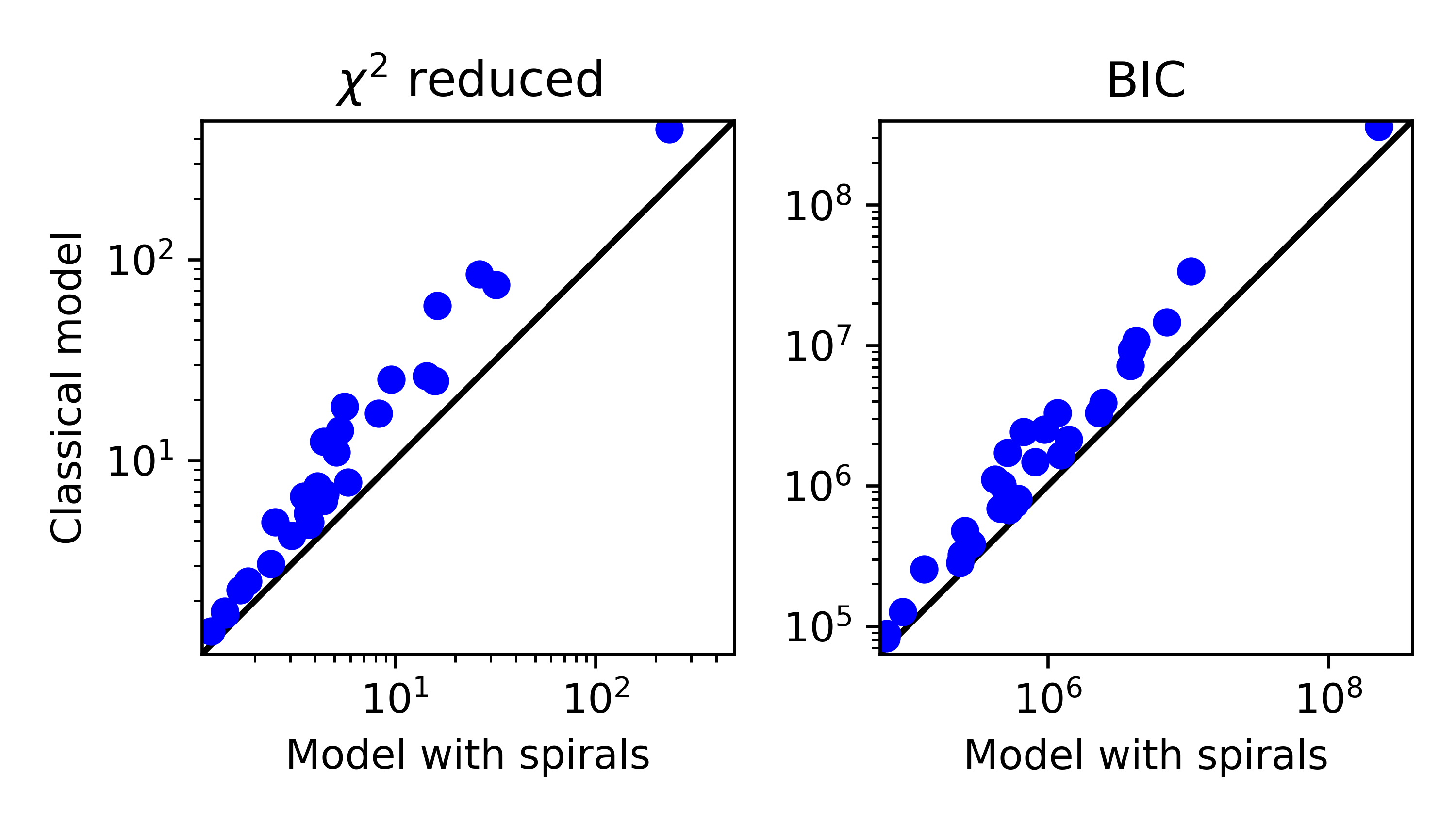}
   \caption{$\chi^2$ (left) and BIC (right) statistic values. Each dot marks corresponding value for two models of galaxy: model with spirals (x axis) and classical one (y axis). We note that chi-square statistic ($\chi^2_\nu$) values seems unrealistically large, with the highest value $\chi^2_\nu > 200$ even for a model with spiral arms of M\,51 (NGC\,5194). However, \citet{Salo2015} also obtained $\chi^2_\nu$ values much higher than unity for some galaxies using exactly the same images and error maps as we used in this study. For example, the bulge+disc decomposition model of M\,51 in \citet{Salo2015} has $\chi^2_\nu \approx$ 1600.
   }
   \label{fig:stats}
\end{figure}

In order to validate our results, we perform some checks. Firstly, we measure the $\chi^2$ statistic for both models of each galaxy. 
Models with spiral arms all have better $\chi^2$ values than classical models, as seen in Fig.~\ref{fig:stats}. However models with a greater number of fitting parameters should generally have better $\chi^2$ value, and to account for this, we use BIC (Bayesian Information Criterion,~\citealt{Bailer-Jones2017}) statistics. BIC considers not only the difference between a model and an image, but also the complexity of the model, giving a penalty for a greater number of parameters. Again, in Fig.~\ref{fig:stats}, one can see that the BIC values also improve for models with spiral arms. This proves that adding spiral arms to our models is justified and does not cause the overfitting problem.

\begin{figure} 
   \centering
   \includegraphics[width=0.95\columnwidth, angle=0]{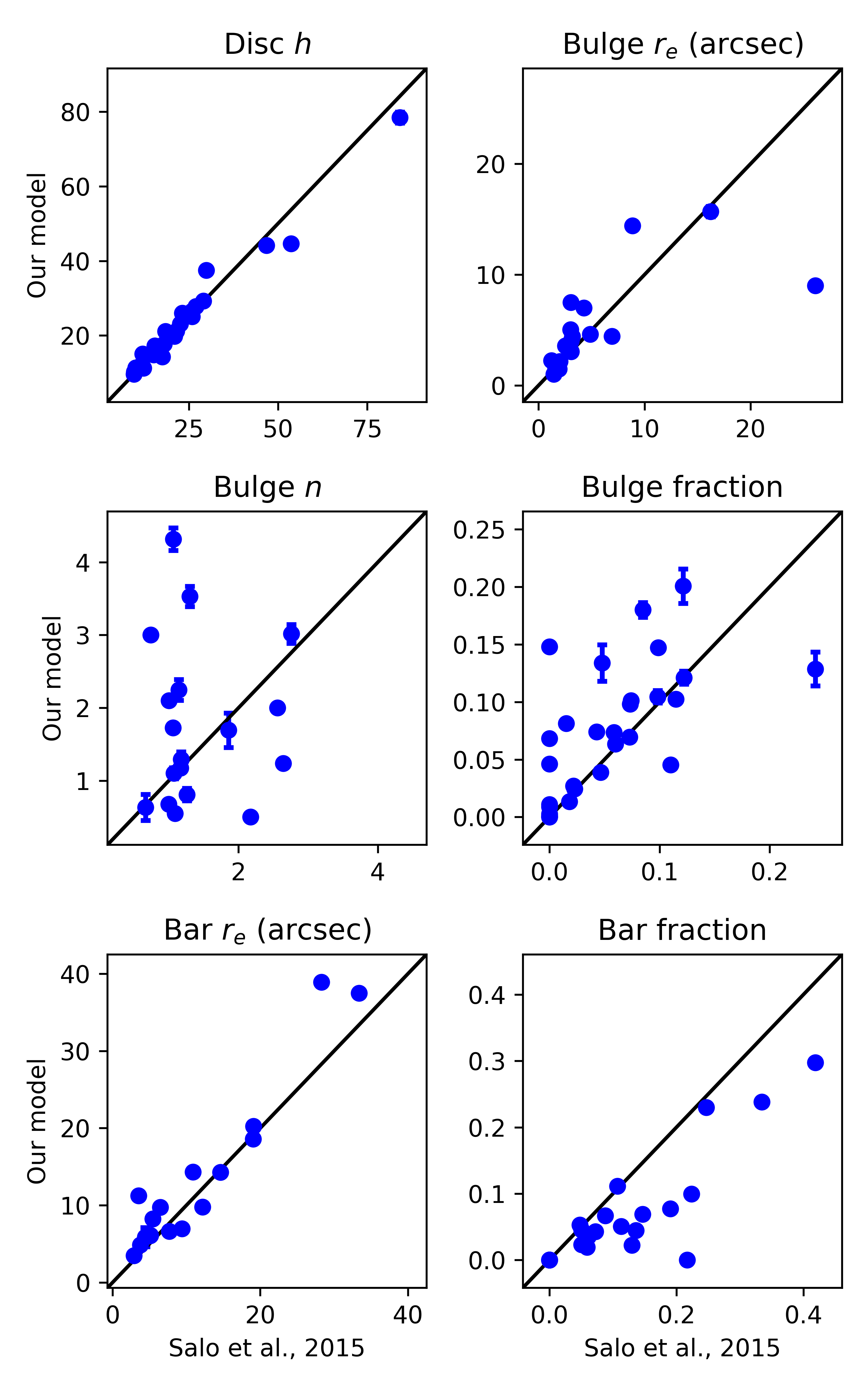}
   \caption{Comparison of different fit parameters derived for each galaxy between~\citet{Salo2015} and classical models obtained in this study.} 
   \label{fig:Salo_comp}
\end{figure}

We can compare our results for classical models with those from~ \citet{Salo2015}, which, in some extent, served as a source (or as a reference, at least) of the initial guess for our models. For some galaxies, we adopted a different set of components (usually more detailed) than the authors of the aforementioned work. Nevertheless, we expect that the general parameters, such as the disc scale length or bulge-to-total ratio, should remain the same. The comparison of several parameters is shown in Fig.~\ref{fig:Salo_comp}. For some parameters (the S{\'e}rsic index $n$ for bulges and $B/T$), we see only a rough correspondence with a large scatter at best. Possibly, bulges in our sample are faint and the addition of different components changes their parameters significantly, in agreement with \citet{Lasker2014}. At the same time, there is a good agreement for other parameters, such as the disc scale length $h$, bulge effective radius $r_\text{e}^\text{bulge}$, and bar effective length $r_\text{e}^\text{bar}$. Interestingly, we notice that bars in~\citet{Salo2015} are systematically stronger than those in our work. Perhaps, this can be attributed to the fact that for most galaxies in our sample we used generalized S{\'e}rsic ellipses instead of a Ferrers function, or due to the inclusion of other components.

\section{The effects of inclusion of spiral arms on the model components}
\label{sec:effect_on_other}

To show the influence of the spiral pattern on the results of decomposition, we compared the parameters of the components in our classical models and models with spiral arms. The difference between them can be interpreted as a systematic error of the parameters estimation of classical decomposition. A comparison of the disc and bulge parameters between the classical models and models with spiral arms is shown in Table.~\ref{tab:galaxies_parameters}.

\begin{table*}
\caption[]{Difference of the parameters between classical models and models with spiral arms}
\begin{centering}
\label{tab:galaxies_parameters}
\begin{tabular}{lccccccccccccc}
\hline\noalign{\smallskip}
Galaxy & \multicolumn{2}{|c|}{Disc} & \multicolumn{6}{c|}{Bulge} & \multicolumn{4}{c|}{Bar} & Components \\
& $h^\text{c} / h^\text{sp}$ & $I_0^\text{c} / I_0^\text{sp}$ & $r_\text{e}^\text{c} / r_\text{e}^\text{sp}$ & $I_\text{e}^\text{c} / I_\text{e}^\text{sp}$ & $n^\text{sp}$ & $n^\text{c}$ & $(B/T)^\text{sp}$ & $(B/T)^\text{c}$ & $r_\text{e}^\text{c} / r_\text{e}^\text{sp}$ & $I_\text{e}^\text{c} / I_\text{e}^\text{sp}$ & $(\text{Bar}/T)^\text{sp}$ & $(\text{Bar}/T)^\text{c}$ &\\
\hline\noalign{\smallskip}
ESO508-024 & 0.96 & 1.23 & 1.00 & 0.44 & 0.52 & 0.72 & 0.03 & 0.01 & 1.50 & 0.49 & 0.08 & 0.11 & DBbar \\
IC0769 & 1.03 & 1.38 & --- & --- & --- & --- & 0.00 & 0.00 & 0.68 & 1.76 & 0.11 & 0.07 & DBbar \\
IC1993 & 1.18 & 1.05 & 0.68 & 1.89 & 1.81 & 1.10 & 0.04 & 0.02 & --- & --- & --- & --- & DB \\
IC2627 & 1.06 & 1.23 & 0.87 & 1.20 & 1.54 & 1.29 & 0.08 & 0.06 & --- & --- & --- & --- & DB \\
IC4237 & 1.00 & 1.14 & --- & --- & --- & --- & 0.01 & 0.01 & 1.00 & 1.00 & 0.04 & 0.05 & DBbarRL \\
NGC0895 & 1.06 & 1.39 & 1.00 & 0.95 & 0.91 & 1.00 & 0.06 & 0.07 & --- & --- & --- & --- & DB \\
NGC0986 & 1.00 & 1.55 & 1.54 & 0.39 & 1.05 & 2.25 & 0.16 & 0.20 & 1.00 & 0.94 & 0.26 & 0.24 & DBbarR \\
NGC2460 & --- & --- & 0.88 & 1.38 & 4.90 & 4.32 & 0.12 & 0.13 & --- & --- & --- & --- & DB \\
NGC3507 & 1.07 & 1.22 & 0.61 & 2.69 & 4.76 & 3.02 & 0.10 & 0.07 & 1.12 & 0.84 & 0.06 & 0.05 & DBbar \\
NGC3596 & 0.67 & 4.32 & 0.41 & 1.79 & 1.22 & 1.24 & 0.22 & 0.07 & --- & --- & --- & --- & DB \\
NGC3683A & 1.01 & 1.29 & --- & --- & --- & --- & 0.00 & 0.00 & 0.67 & 1.32 & 0.04 & 0.02 & DBbar \\
NGC3684 & 0.87 & 1.55 & 3.73 & 0.10 & 0.61 & 3.53 & 0.03 & 0.08 & 0.84 & 0.58 & 0.10 & 0.02 & DBbar \\
NGC3686 & 1.02 & 1.19 & 0.90 & 1.27 & 2.22 & 1.69 & 0.03 & 0.03 & 1.05 & 0.95 & 0.03 & 0.02 & DBbar \\
NGC3687 & 0.92 & 1.74 & 0.67 & 2.67 & 3.23 & 1.17 & 0.15 & 0.10 & 1.06 & 0.72 & 0.07 & 0.05 & DBbar \\
NGC4067 & 1.14 & 1.29 & 1.00 & 0.76 & 2.69 & 3.00 & 0.17 & 0.13 & 0.94 & 1.10 & 0.10 & 0.10 & DBbar \\
NGC4165 & 1.00 & 1.24 & 0.73 & 1.35 & 1.04 & 0.63 & 0.06 & 0.04 & --- & --- & --- & --- & DB \\
NGC4314 & 0.67 & 6.98 & 0.80 & 1.36 & 1.10 & 0.55 & 0.14 & 0.10 & 1.00 & 0.84 & 0.30 & 0.30 & DBbarRL \\
NGC4548 & --- & --- & 0.97 & 1.10 & 2.42 & 2.00 & 0.15 & 0.15 & 1.08 & 0.86 & 0.22 & 0.23 & DBbar \\
NGC4680 & 0.91 & 1.71 & --- & --- & --- & --- & 0.05 & 0.05 & 0.93 & 0.85 & 0.08 & 0.04 & DBbar \\
NGC4902 & 0.93 & 2.21 & 0.82 & 1.11 & 0.91 & 0.81 & 0.10 & 0.07 & 1.00 & 1.09 & 0.10 & 0.07 & DBbar \\
NGC5194 & 0.94 & 2.44 & 0.70 & 1.41 & 1.07 & 0.67 & 0.23 & 0.12 & --- & --- & --- & --- & DB \\
NGC5240 & 1.10 & 1.25 & 0.45 & 3.61 & 1.15 & 0.50 & 0.02 & 0.01 & --- & --- & --- & --- & DB \\
NGC5247 & --- & --- & 1.34 & 0.60 & 1.29 & 1.72 & 0.12 & 0.18 & --- & --- & --- & --- & DB \\
NGC5364 & 1.07 & 1.24 & 1.11 & 0.59 & 1.37 & 2.10 & 0.05 & 0.05 & --- & --- & --- & --- & DB \\
NGC5427 & 0.93 & 1.92 & 0.79 & 1.48 & 0.50 & 0.50 & 0.11 & 0.10 & --- & --- & --- & --- & DB \\
NGC7167 & 1.04 & 0.99 & 1.73 & 0.91 & 3.17 & 1.84 & 0.09 & 0.15 & --- & --- & --- & --- & DB \\
NGC7661 & 1.09 & 1.07 & --- & --- & --- & --- & --- & --- & 0.90 & 1.06 & 0.06 & 0.04 & Dbar \\
NGC7798 & 1.04 & 1.42 & --- & --- & --- & --- & 0.11 & 0.10 & 1.00 & 1.26 & 0.07 & 0.04 & DBbar \\
PGC028380 & 0.94 & 1.15 & --- & --- & --- & --- & --- & --- & 1.03 & 0.90 & 0.08 & 0.08 & Dbar \\
\hline\noalign{\smallskip}
\end{tabular}\\
\end{centering}
\textit{Notes}: index $^\text{c}$ stands for ``classical'' model parameter, $^\text{sp}$ stands for a model with spiral arms. Letter combination in ``Components'' describes the set of components except spiral arms which were used to fit each galaxy: D --- disc, B --- bulge, bar --- bar, R --- ring, and L --- lens. Different functions were used to fit these galaxy components. For example, exponential or broken exponential functions were used to fit the disc, and a point source function was sometimes used for the bulge when it was too small to be resolved. Dashes in the table indicate that the corresponding parameter is undefined for this galaxy, as in the case of broken exponential discs in NGC~2460, NGC~4548, and NGC~5247, or bulges modelled as a point source (only $B/T$ was determined in such cases), or the corresponding component does not present in the galaxy.
\end{table*}

\subsection{The disc parameters}
\label{sec:disc_params}

\begin{figure} 
   \centering
   \includegraphics[width=0.95\columnwidth, angle=0]{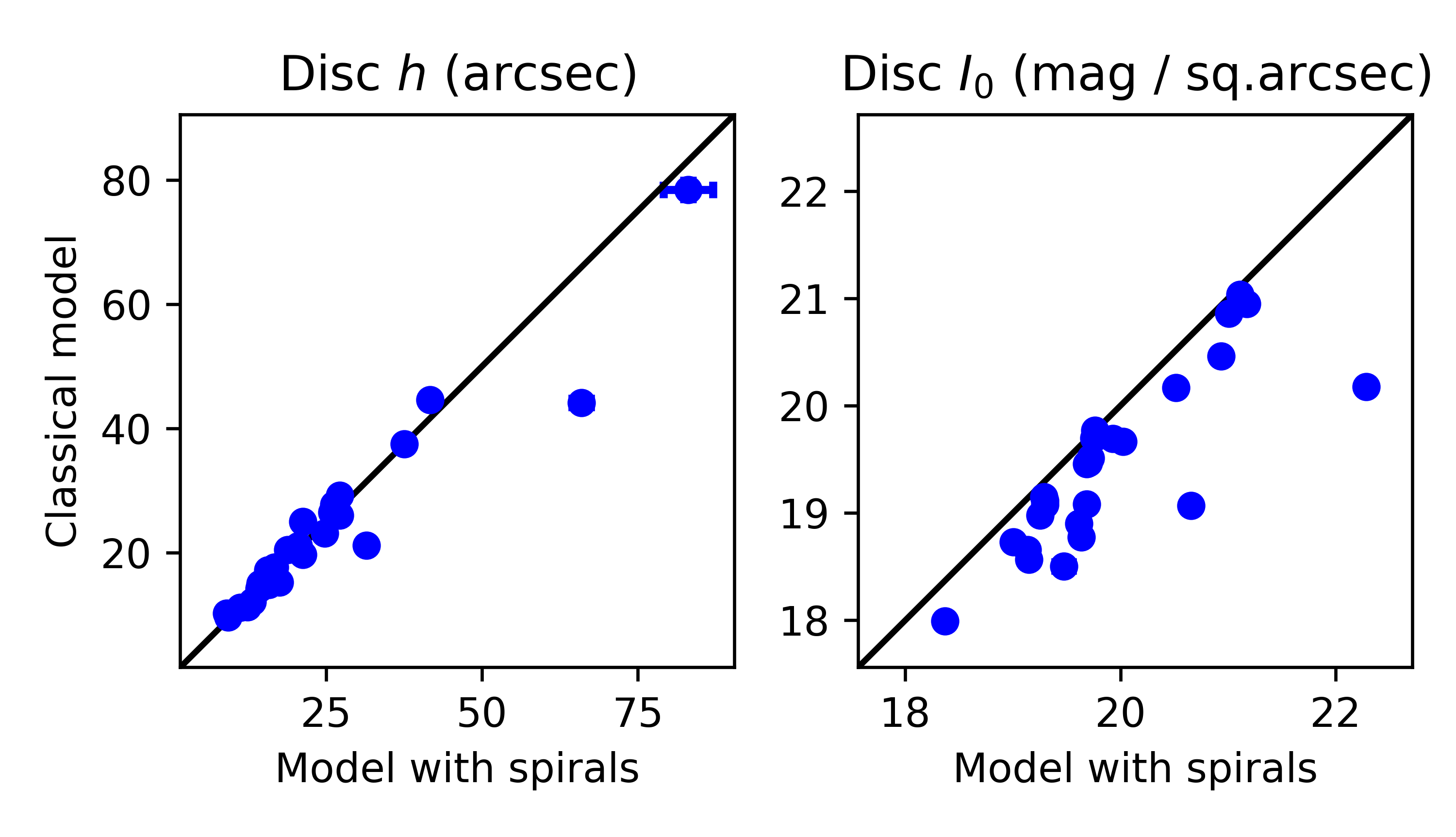}
   \caption{Difference of the disc parameters between classical models and models with spiral arms. The radial scale length $h$ is shown on the left, the central surface brightness $I_0$ is displayed on the right.} 
   \label{fig:disc_diff}
\end{figure}

The central surface brightness of discs ($I_0$) appears to be higher in classical models than in models with spiral arms, see Fig.~\ref{fig:disc_diff} and Table~\ref{tab:galaxies_parameters}. This is expected because the spiral arms are disc features, and if they are included, they take some part of the disc light. The mean difference for our sample is 0.5 magnitudes. The radial scale of discs in our sample ($h$) does not change systematically between the classical models and models with spiral arms, so the difference for individual galaxies is small. When spiral arms are added to the model, $h$ may increase or decrease, but this change does not exceed 10\% in most of the cases.

\par

\citet{Gao2017} found that the radial scale length of the disc increases when spiral arms are added to the model, reasoning that spiral arms have a truncation at large radii. Following this, when one uses a pure disc model to fit a disc with spiral arms, the disc scale length becomes smaller than its actual value to account for this truncation. However, spiral arms usually do not emerge from the centers of galaxies, so spiral structure should have a brightness drop in the center which should compensate to a some extent the effect of outer truncation. \citet{Gao2017} used a spiral arms model from {\small{GALFIT}}~\citep{Peng2010} (which actually represents an azimuthally-distorted disc) separately from the axisymmetric disc component. This means that their model of spiral structure does not have a drop of surface brightness in the center. In other words, their model of spiral arms has an excess of light distribution in the center, so the disc model in this central region is underestimated which makes the axisymmetric disc profile less steep. Another decomposition with a simple model of spiral arms was carried out in~\citet{Lingard2020}. For disc size and ellipticity, they did not found any significant difference between their models with spiral arms and without them.

\subsection{The bulge parameters}
\label{sec:bulge_params}
\begin{figure} 
   \centering
   \includegraphics[width=0.95\columnwidth, angle=0]{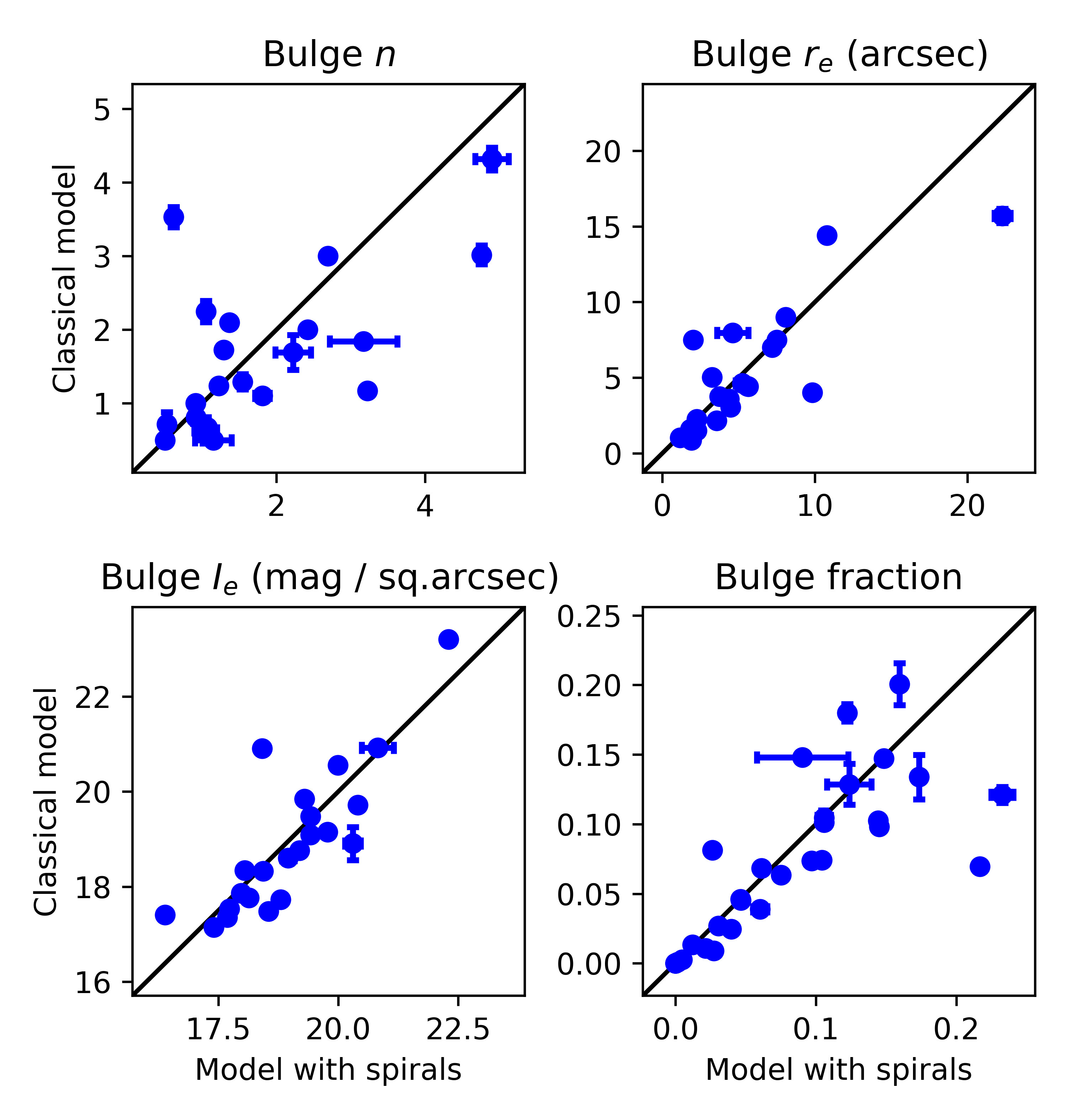}
   \caption{Difference of the bulge parameters between classical models and models with spiral arms. Upper left: S{\'e}rsic index $n$; upper right: effective radius $r_\text{e}^\text{bulge}$; lower left: effective surface brightness $I_\text{e}^\text{bulge}$; lower right: bulge-to-total luminosity ratio.} 
   \label{fig:bulge_diff}
\end{figure}

The systematic difference of the bulge parameters (effective radius $r_\text{e}^\text{bulge}$, effective surface brightness $I_\text{e}^\text{bulge}$, S{\'e}rsic index $n$, or bulge-to-total luminosity ratio $B/T$) between our classical models and models with spiral arms is noticeable only for the sample as a whole (see Fig.~\ref{fig:bulge_diff} and Table~\ref{tab:galaxies_parameters}). When spiral arms are added, the mean changes of parameters are the following: $r_\text{e}^\text{bulge}$ increases by 20\%, $I_\text{e}^\text{bulge}$ decreases by 5\%, $n$ increases by 26\% and $B/T$ increases by 33\%. However, the scatter around these values is large for individual galaxies. For example, the $B/T$ and $n$ parameters can increase or decrease by factor of two when spiral arms added to the model. The explanation of the average trend in the models is as follows. When spiral arms are included in the model, the measured central surface brightness of the disc becomes lower. The spiral arms are not present in the center, so the decrease of the surface brightness in the center should be compensated by some other components, such as the bar or the bulge. The shape of the distribution of the ``remaining'' luminosity density, to be fitted with the bulge and other components, changes and the parameters of these components may be different. The brightness in the very center of the bulge does not change much because in this point it is already much higher than disc brightness. In the same time, extended ``wings'' may appear in the outer parts of a bulge which require an increased $n$. Overall, the bulge becomes more extended and more luminous, which leads to the increase of $r_\text{e}^\text{bulge}$ and $B/T$. The decrease of $I_\text{e}^\text{bulge}$ may seem to contradict to these reasons, but, as the bulge becomes more extended, its half-light radius shifts to the outer fainter parts, which leads to the decrease of $I_\text{e}^\text{bulge}$, while the central brightness of the bulge remains nearly constant with an increased $n$.

\par

Therefore, we conclude that neglecting spiral arms in decomposition may lead to significant errors and biases in the estimated bulge parameters for individual galaxies.
 \citet{Gao2017} reported that adding spiral arms in their model changes the bulge parameters insignificantly, however their sample contained only 6 spiral galaxies which seems to be not enough to draw robust statistical conclusions considering the large scatter in our results. As discussed in Sec.~\ref{sec:disc_params}, we can explain this discrepancy by the fact that the spiral arms model in~\citet{Gao2017} does not have a drop of surface brightness in the center. When spiral arms are added to the model, the decrease of the central surface brightness of the disc is compensated by the model of spiral arms. Therefore, the ``remaining'' surface brightness, which is fitted by a bulge, remains the same as in the model without spirals, and the bulge parameters are not expected to change. \citet{Sonnenfeld2022} performed a S{\'e}rsic profile fitting for artificial images of galaxies with spiral arms and found that the half-light radius of galaxy and the total flux is overestimated by 30\% and 15\%, respectively, for the spiral structure contributing 10\% of total light but for disc-dominated galaxies this bias is much smaller. In our sample, galaxies have small $B/T$ and are disc-dominated. \citet{Lingard2020} found that the bulge-to-total ratio is higher in their models with spiral arms than in simple S{\'e}rsic bulge + exponential disc models.

\subsection{The bar parameters}
\label{sec:bar_params}
\begin{figure} 
   \centering
   \includegraphics[width=0.95\columnwidth, angle=0]{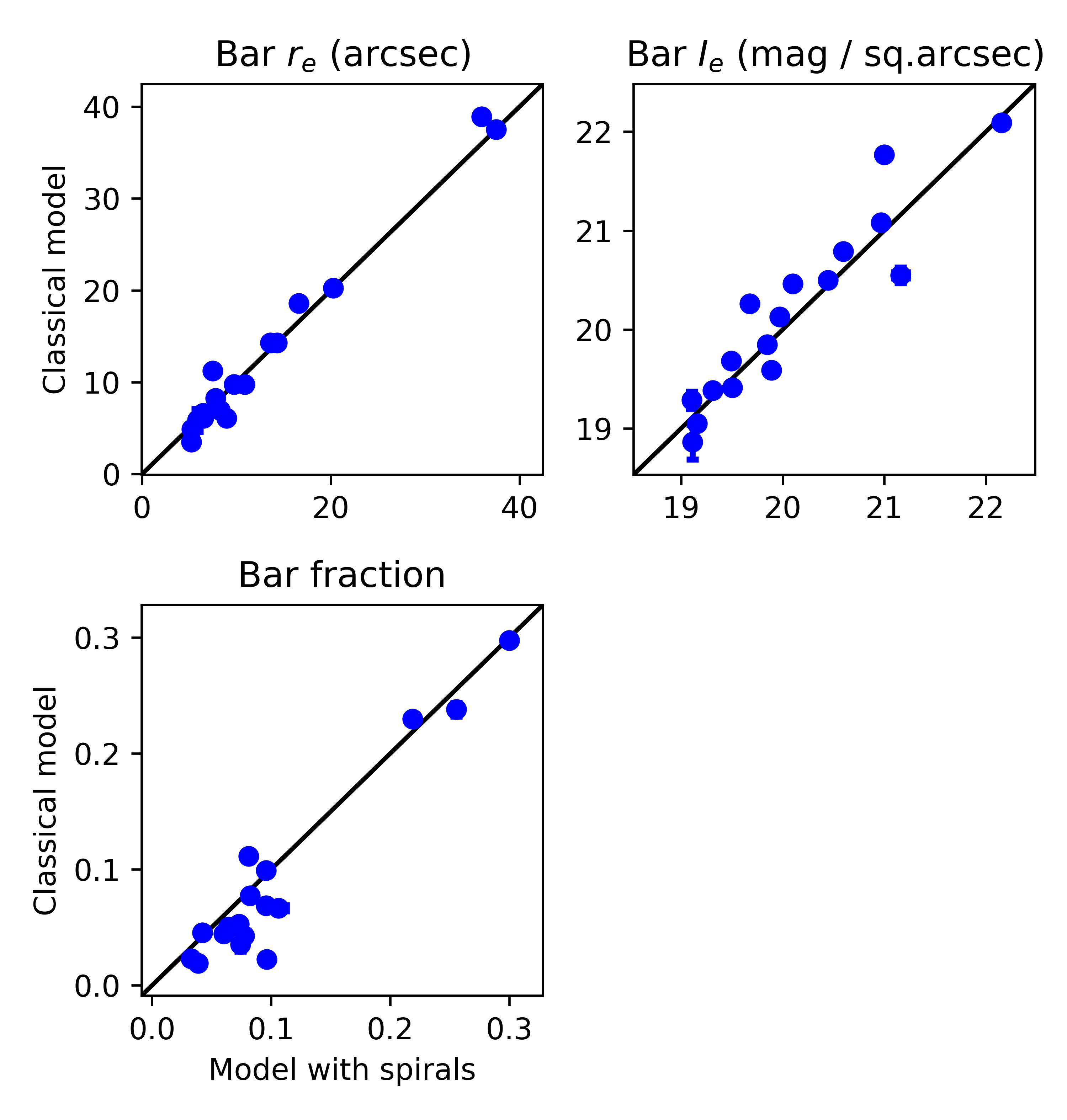}
   \caption{Difference of bar parameters between classical models and models with spiral arms. Upper left: effective radius $r_\text{e}^\text{bar}$; upper right: effective surface brightness $I_\text{e}^\text{bar}$; lower left: bar-to-total luminosity ratio.} 
   \label{fig:bar_diff}
\end{figure}

Among 29 galaxies from our sample, 17 are barred, so the effects from inclusion of spiral arms on the bar parameters are also interesting to investigate. In Fig.~\ref{fig:bar_diff} and Table~\ref{tab:galaxies_parameters}, we present a comparison for the bar size (expressed in terms of the effective radius along the major axis $r_\text{e}^\text{bar}$), bar effective surface brightness $I_\text{e}^\text{bar}$, and bar-to-total luminosity ratio $\text{Bar}/T$. When spiral arms are added to the model, the mean $r_\text{e}^\text{bar}$ increases by 5\%, the mean $\text{Bar}/T$ increases by 49\% with the most significant changes occur when $\text{Bar}/T$ is small, and the mean $I_\text{e}^\text{bar}$ increases by 7\%. Just as for bulge parameters, the scatter is large. Therefore, $\text{Bar}/T$ increases stronger than the $B/T$ when spiral arms are added but the change of $r_\text{e}^\text{bar}$ is smaller than the change of $r_\text{e}^\text{bulge}$.

We suggest that such a difference may be caused by the fact that the bulges in our sample are small, concentrated, and have extended ``wings'' on their surface brightness profiles (a high S{\'e}rsic index $n$), while the bars usually are larger and have flatter brightness profiles (since $n$ is usually smaller than $1$ for them) and lower effective surface brightnesses. This means that a slight variation of the disc brightness in our models, caused by the inclusion of spiral arms, affects the bar and bulge parameters differently. For bulges, the mentioned variation affects primarily the faint ``wings'', and the best-fitting bulge model turns out to have a much brighter (or fainter) periphery. This means that $n$ changes significantly, and other parameters are also changed to retain the surface brightness in the central part. Bars, conversely, usually have well-defined borders, and their general shape remains the same. However, even a small increase of the disc brightness can decrease the overall bar luminosity substantially due to the bar large size and surface brightness lower than the bulge has. The bar effective surface brightness also varies significantly, in consistency with this interpretation.

\section{Spiral arms parameters}
\label{sec:spiral_parameters}

From our decomposition, we have obtained a parametric description of the galaxies from our sample. We can now determine different properties of the spiral arms, including their width, pitch angle, and how these parameters vary with radius.

\begin{table*}
\caption{Parameters of the spiral arms in our sample galaxies}
\begin{centering}
\label{tab:spiral_parameters}
\begin{tabular}{lccccccccc}
    \hline\noalign{\smallskip}
    Galaxy & $S/T$ & $\langle h_\text{s} \rangle / h$ & $\langle w \rangle / h$ & $\sigma_\text{w} / \langle w \rangle$ & $\langle A \rangle$ & $\langle \upmu \rangle$ & $\sigma_\upmu$ & $\Delta \upmu$ & $N_\text{arms}$\\
    & & & & & & deg & deg & deg & \\
    & (1) & (2) & (3) & (4) & (5) & (6) & (7) & (8) & (9) \\
    \hline\noalign{\smallskip}
ESO508-024 & 0.08 & 0.09 & 0.24 & 0.02 & 0.23 & 18.4 & 1.6 & 5.6 & 2 \\
IC0769 & 0.21 & 1.47 & 0.52 & 0.02 & 0.00 & 13.9 & 0.4 & 2.4 & 2 \\
IC1993 & 0.21 & 0.86 & 0.46 & 0.25 & 0.43 & 11.9 & 0.8 & 8.3 & 3 \\
IC2627 & 0.24 & 2.74 & 0.64 & 0.50 & $-$0.04 & 28.0 & 9.2 & 12.8 & 3 \\
IC4237 & 0.25 & 3.61 & 0.52 & 0.24 & 0.25 & 12.9 & 5.2 & 2.1 & 3 \\
NGC0895 & 0.25 & 2.28 & 0.29 & 0.02 & $-$0.30 & 18.4 & 0.5 & 6.7 & 2 \\
NGC0986 & 0.18 & 0.34 & 0.52 & 0.09 & $-$0.48 & 13.9 & 1.4 & 18.4 & 2 \\
NGC2460 & 0.29 & --- & --- & 0.76 & $-$0.21 & 11.0 & 12.0 & 7.6 & 5 \\
NGC3507 & 0.21 & 0.68 & 0.57 & 0.12 & $-$0.02 & 10.6 & 1.3 & 5.9 & 2 \\
NGC3596 & 0.39 & 0.26 & 0.44 & 0.12 & 0.17 & 14.0 & 2.8 & 5.1 & 2 \\
NGC3683A & 0.16 & 1.88 & 0.54 & 0.60 & $-$0.20 & 17.5 & 4.2 & 8.6 & 4 \\
NGC3684 & 0.14 & 0.64 & 0.50 & 0.27 & $-$0.13 & 24.3 & 9.7 & 9.0 & 3 \\
NGC3686 & 0.18 & 1.07 & 0.55 & 0.20 & 0.02 & 18.2 & 6.9 & 5.2 & 2 \\
NGC3687 & 0.22 & 0.77 & 0.75 & 0.56 & $-$0.16 & 11.3 & 4.6 & 6.9 & 3 \\
NGC4067 & 0.20 & 1.68 & 0.80 & 0.28 & $-$0.18 & 8.6 & 6.6 & 9.2 & 3 \\
NGC4165 & 0.17 & 0.89 & 0.61 & 0.01 & $-$0.12 & 11.2 & 2.8 & 4.7 & 2 \\
NGC4314 & 0.14 & 0.31 & 0.78 & 0.02 & $-$0.27 & 8.0 & 1.2 & 21.3 & 2 \\
NGC4548 & 0.14 & --- & --- & 0.48 & $-$0.21 & 13.0 & 8.1 & 15.1 & 4 \\
NGC4680 & 0.27 & 0.80 & 0.41 & 0.25 & $-$0.16 & 15.6 & 3.4 & 8.4 & 2 \\
NGC4902 & 0.38 & 0.61 & 0.66 & 0.15 & $-$0.37 & 14.9 & 5.4 & 2.9 & 3 \\
NGC5194 & 0.46 & 0.41 & 0.52 & 0.14 & $-$0.21 & 15.2 & 0.5 & 4.4 & 2 \\
NGC5240 & 0.24 & 1.29 & 0.45 & 0.22 & 0.31 & 25.4 & 4.5 & 9.1 & 3 \\
NGC5247 & 0.32 & --- & --- & 0.10 & $-$0.30 & 29.4 & 5.3 & 5.4 & 3 \\
NGC5364 & 0.26 & 0.94 & 0.45 & 0.13 & 0.14 & 11.8 & 2.5 & 5.4 & 2 \\
NGC5427 & 0.36 & 1.02 & 0.48 & 0.39 & $-$0.42 & 20.3 & 6.5 & 7.4 & 4 \\
NGC7167 & 0.13 & 0.77 & 0.34 & 0.01 & $-$0.30 & 21.0 & 2.5 & 9.8 & 2 \\
NGC7661 & 0.14 & 3.45 & 1.18 & 0.14 & $-$0.12 & 27.8 & 4.2 & 12.3 & 2 \\
NGC7798 & 0.19 & 3.73 & 0.33 & 0.41 & 0.60 & 11.0 & 7.7 & 4.8 & 3 \\
PGC028380 & 0.12 & 7.58 & 0.55 & 0.29 & 0.65 & 21.5 & 1.2 & 6.8 & 2 \\
    \hline\noalign{\smallskip}
\end{tabular}\\
\end{centering}
(1) Spiral-to-total ratio\\
(2) Mean spiral arm exponential scale relative to the disc exponential scale\\
(3) Mean spiral arm width relative to the disc exponential scale\\
(4) Standard deviation of the spiral arm width relative to its mean value\\
(5) Mean asymmetry of the spiral arms\\
(6) Mean pitch angle of the spiral arms\\
(7) Standard deviation of averaged pitch angles of arms in galaxy\\
(8) Average value of variations in individual arms of galaxy\\
(9) Number of spiral arms\\
\end{table*}

\subsection{The fraction of the spiral arms in the total galaxy luminosity}

\begin{figure}
\centering
\includegraphics[width=0.95\columnwidth, angle=0]{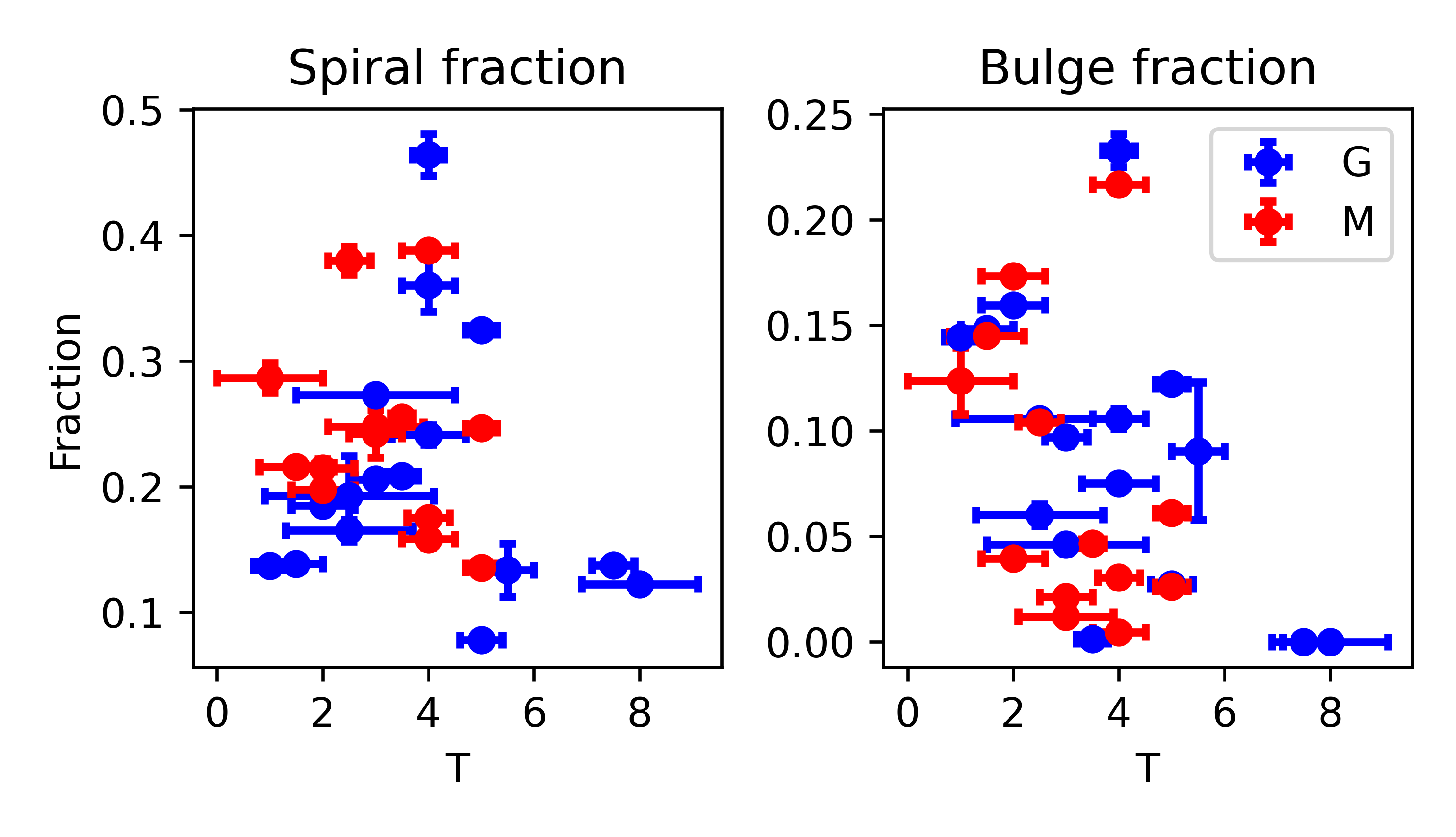}
\caption{Relation between the Hubble type (from~\citealt{Buta2015}) and the fraction of spiral arms.}
\label{fig:T_fracs}
\end{figure}

We first inspect the relative fraction of the spiral arms in the total galaxy luminosity ($S/T$) and the bulge fraction ($B/T$) from our models versus Hubble types adopted from~\citet{Buta2015}. For most galaxies in our sample, $S/T$ is found between 10\% and 25\%. However in exceptional cases, the fraction of spiral arms may exceed 45\%, as seen in Table~\ref{tab:spiral_parameters}. The highest value of $S/T$ is achieved for intermediate-type spirals, see Fig.~\ref{fig:T_fracs}, and $B/T$ is higher in early-type spirals, as expected. However, intermediate-type spirals with a low $S/T$ are also present. We note that early-type spiral galaxies are located near the Hubble stage $T = 0$ which marks a transition to lenticular galaxies without spiral arms. Early-type spirals tend to have a low gas mass fraction and low star formation rate, and, therefore, they cannot have luminous starforming spiral arms~\citep{Roberts1994}. Very late-type spirals, located near the transition to irregular galaxies, also lack an ordered, well-defined spiral structure. Such galaxies tend to have multiple flocculent arms, which are hard to fit with our model, and are often blended with the disc component. \citet{Savchenko2020} found that the fraction of spiral structure is higher in Sc-type galaxies than in Sa-type ones, in agreement with our result, and the similar overall values of $S/T$.

\begin{figure*}
\centering
\includegraphics[width=1.95\columnwidth, angle=0]{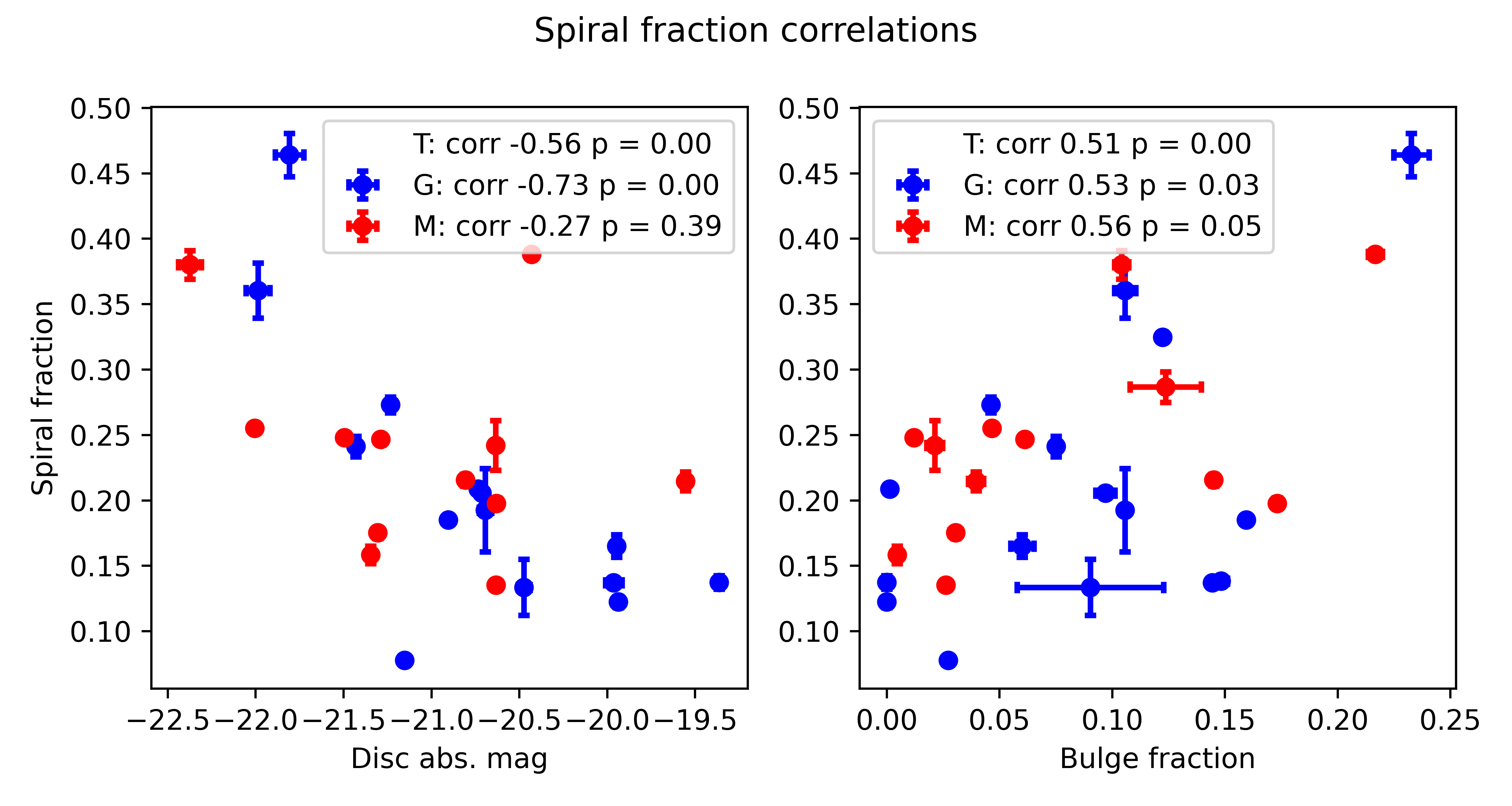}
\caption{Correlations between $S/T$ and $B/T$ (left) and between $S/T$ and disc absolute magnitude (right).}
\label{fig:ST_corrs}
\end{figure*}

In Fig.~\ref{fig:ST_corrs}, we display correlations between $S/T$ and $B/T$ and between $S/T$ and disc absolute magnitude. For both types of spirals, grand-design and multi-armed, $S/T$ is higher in galaxies with a higher $B/T$. The same relation between $S/T$ and $B/T$ was found in~\citet{Bittner2017} for galaxies with a low $B/T$, and the galaxies in our study have low $B/T$ compared to the sample in the mentioned work. For grand-design galaxies, we also find that $S/T$ is higher in galaxies having more luminous discs. This finding matches the fact that irregular galaxies have faint discs and, at the same time, demonstrate no clear spiral structure.

\subsection{Pitch angles}

Our method allows us to measure not only pitch angles of individual spiral arms, but also their variations along a single arm. In contrast, widely-used Fourier-based methods provide only the pitch angle averaged over the entire galaxy. In Table~\ref{tab:spiral_parameters}, we show the following values: the average pitch angle for each galaxy $\langle \upmu \rangle$, standard deviation of the averaged individual pitch angles for all arms $\sigma_\upmu$, and the average value of pitch angle variations in the galaxy arms $\Delta \upmu$. 
Defined this way, $\sigma_\upmu$ describes the difference between all spiral arms in a galaxy, whereas $\Delta \upmu$ describes the variability of the pitch angles. For example, galaxies NGC~986 and NGC~4314 have small $\sigma_\upmu$ but large $\Delta \upmu$. Indeed, both galaxies have two very symmetric arms, albeit their shapes are far from logarithmic spirals with constant pitch angles. Instead, they form pseudorings and their pitch angles turn negative at some point (see on-line materials).

\par

Both $\sigma_\upmu$ and $\Delta \upmu$ are far from zero in many cases. In our sample, the average value of $\sigma_\upmu$ is $4.2^\circ$ and it is $8.0^\circ$ for $\Delta \upmu$. In some cases, $\sigma_\upmu / \upmu$ exceeds $1/2$. In other words, the pitch angles of spiral arms in a single galaxy may vary significantly and the average value of pitch angle is not sufficient to characterize the galaxy spiral structure. Moreover, even the averaged values of individual spiral arms do not seem to be a reliable measure because spiral arms usually have variable pitch angles, in agreement with~\citet{Lingard2021} and again with~\citet{Savchenko2013}.

\par

\begin{figure} 
   \centering
   \includegraphics[width=0.95\columnwidth, angle=0]{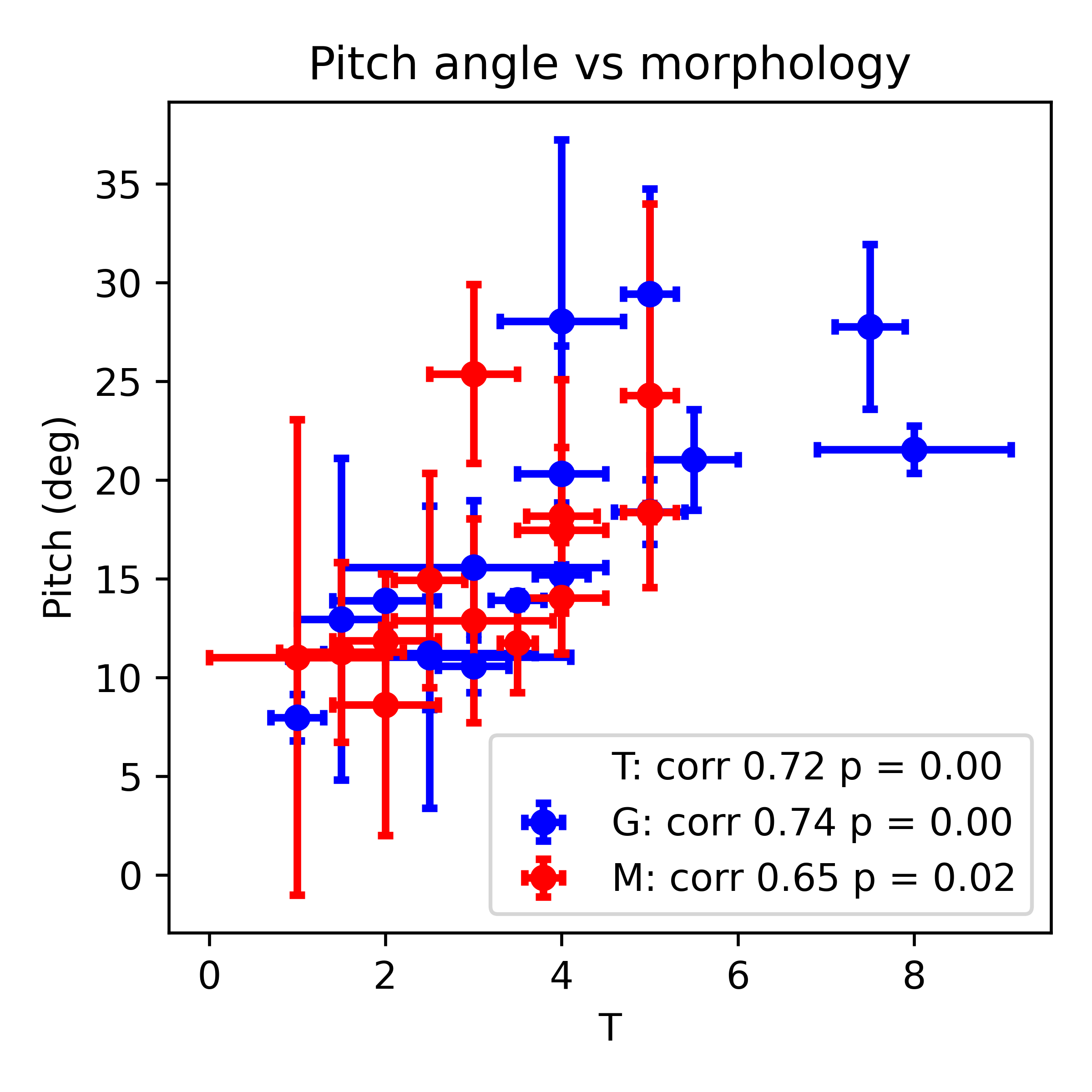}
   \caption{Dependence of the mean pitch angle on morphological type. G --- grand-design galaxies, M --- multi-armed.}
   \label{fig:T-Pitch}
\end{figure}

We confirm a strong relationship between the spiral arm pitch angle and the Hubble type of a galaxy (Fig.~\ref{fig:T-Pitch}). This is a well-established relation which has been reported since~\citet{Kennicutt1981} because the spiral arms' pitch angle is one of the criteria of the Hubble classification, and this result also validates our method. Figure 13 in~\citet{Savchenko2020} shows a correlation between pitch angle and Hubble type based on results from different studies. Although the general trend does exist, the scatter of this correlation is very large, which is also seen in~\citet{Diaz-Garcia2019}. The scatter of our correlation in Fig.~\ref{fig:T-Pitch} is less pronounced, perhaps due to the more accurate estimation of the pitch angle thanks to our sophisticated modeling of the galaxy structure.  

\begin{figure*}
\centering
\includegraphics[width=1.95\columnwidth, angle=0]{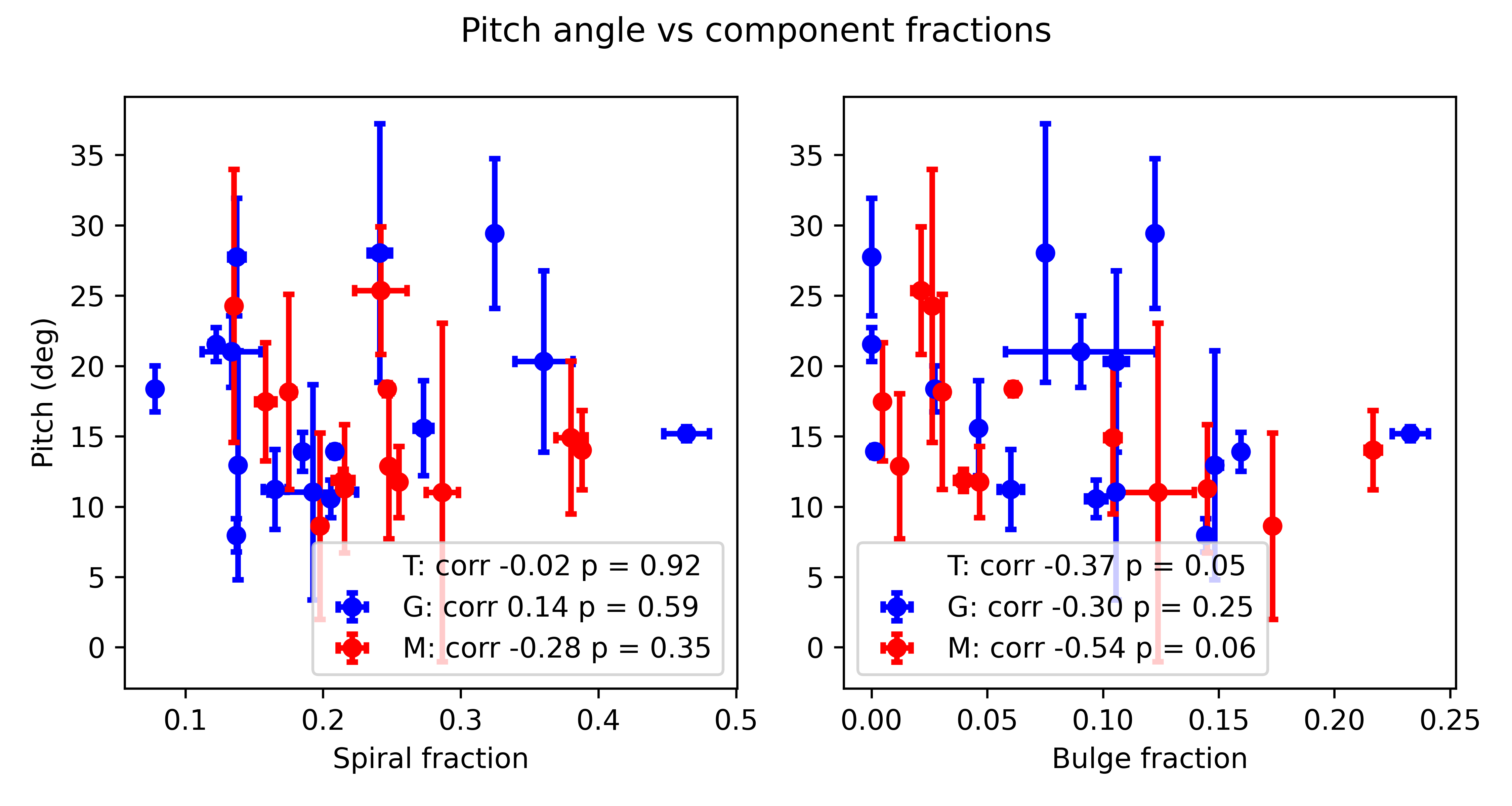}
\caption{Diagrams showing pitch angle versus $S/T$ (left) and $B/T$ (right).}
\label{fig:pitch_corrs}
\end{figure*}

For pitch angles, we do not find any significant correlation with $S/T$, see Fig.~\ref{fig:pitch_corrs}. There are some theoretical studies considering a dependence between these quantities. For example, \citet{Hamilton2023} predicts a positive correlation between the amplitude of spiral structure and its pitch angle. However, \citet{Perez-Villegas2015} predicts a negative correlation between spiral arm mass and pitch angle. Observations show a negative correlation between these parameters \citep{Diaz-Garcia2019}.

\par

We find a weak anticorrelation between pitch angle and $B/T$, see Fig.~\ref{fig:pitch_corrs}. In other words, galaxies with more prominent bulges have more tightly wound spiral arms. This is in agreement with, for example, \citet{Yu2019}. Such a relation is predicted by the density wave theory. However, the very large scatter in this relation observed in \citet{Yu2019}, as well as in this study, is not consistent with the density wave theory~\citep{Masters2020}. In~\citet{Font2019}, a dependence between disc mass fraction and pitch angle of spiral arms was examined for barred galaxies. They did not find any correlation between these parameters, albeit there are no galaxies with low disc mass fraction (and, therefore, with a massive bulge) and a high pitch angle at the same time. \citet{Kendall2015} found no relation between pitch angle and concentration parameter, which can be considered as a proxy of $B/T$. Finally, \citet{Davis2015} reported that the pitch angle decreases with increasing bulge mass.

\begin{figure*}
\centering
\includegraphics[width=1.95\columnwidth, angle=0]{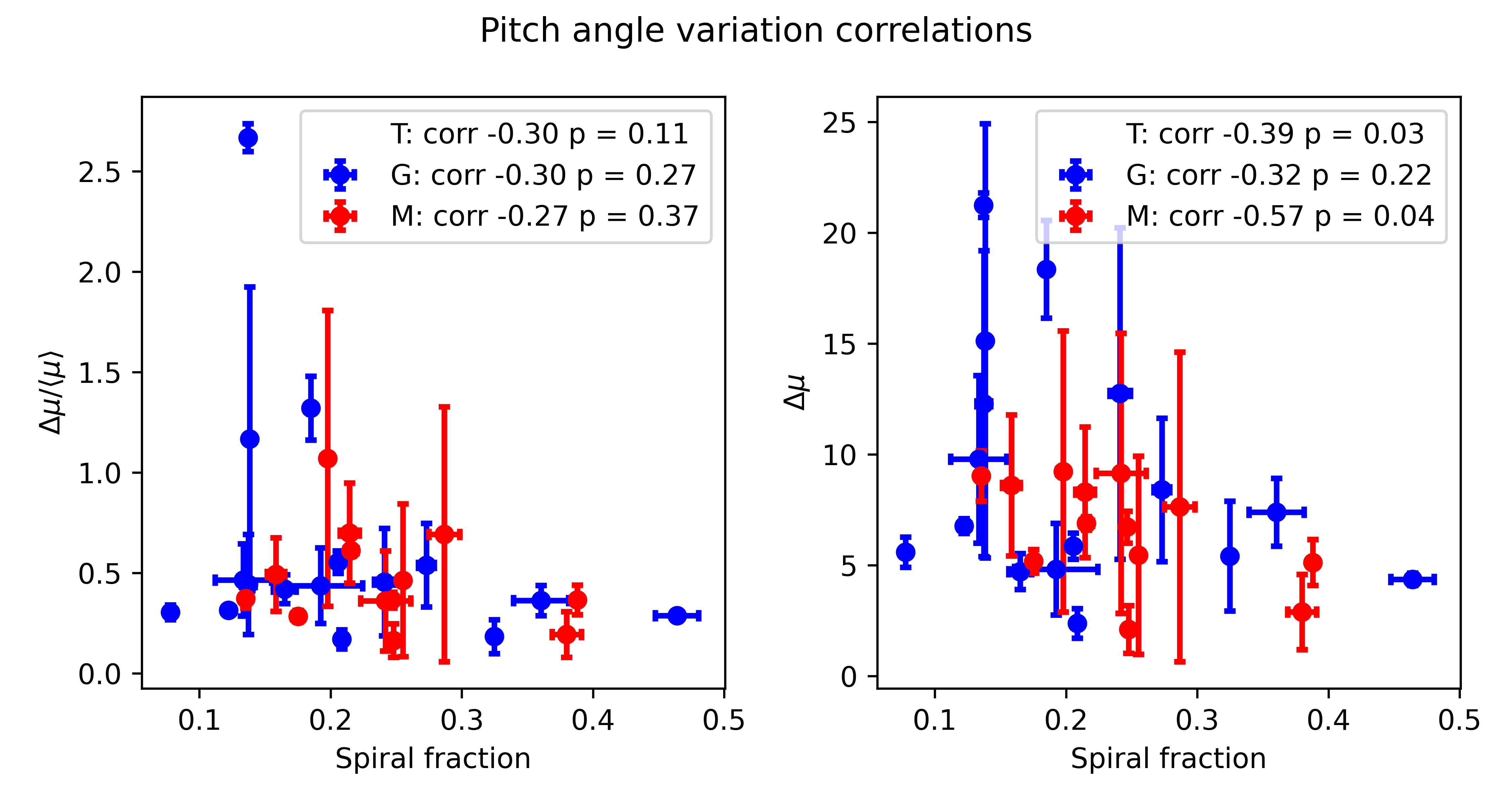}
\caption{Diagrams showing correlations between pitch angle variation $\Delta \upmu$ or $\Delta \upmu / \langle \upmu \rangle$ with spiral-to-total ratio $S/T$ or Hubble type $T$.}
\label{fig:pitch_var_corrs}
\end{figure*}

We also investigate a connection between $\Delta \upmu / \langle \upmu \rangle$ and spiral-to-total luminosity ratio $S/T$. In Fig.~\ref{fig:pitch_var_corrs}, we see a fairly weak anticorrelation between these two quantities, but we note that $\Delta \upmu / \langle \upmu \rangle$ is always small for galaxies with a high $S/T$. In other words, galaxies with a strong spiral structure have spiral arms close to logarithmic (Fig.~\ref{fig:pitch_var_corrs}). \citet{Savchenko2020} found the same dependence. Assuming that Hubble type $T$ is connected with both $\langle \upmu \rangle$ and $S/T$ (see Fig.~\ref{fig:T-Pitch} and~\ref{fig:T_fracs}, respectively), one can think that this relation arises purely from these two relations. However we notice an even stronger relation between $\Delta \upmu$ and $S/T$, which proves that stronger spiral arms indeed have shapes closer to logarithmic spirals.

\subsection{Spiral arm width}

We define the width of a spiral arm as the FWHM of a radial slice in the middle of the arm, i.e. at $\varphi = \varphi_\text{end} / 2$. We note that $w_\text{e}^\text{in}$ and $w_\text{e}^\text{out}$ in eq.~\ref{eq:I_bot} are connected with the width itself not in a simple way, and $w_\text{e}^\text{in/out}$ itself is a value of limited usefulness. The S{\'e}rsic profile implies a symmetric 2D distribution of light, and, in its definition, the effective radius $r_\text{eff}$ is a radius enclosing half of the galaxy flux. 
The denominator in eq.~\ref{eq:I_bot} ($w_\varphi^\text{in/out} = \sqrt{(w_\text{e}^\text{in/out})^2 + \left(\varphi \times \xi \right)^3}$ for short) is similar to $r_\text{eff}$ in its form, but the S{\'e}rsic-like distribution of light in our model is only one-dimensional. If we consider the band of inner/outer half-width of $w_\varphi^\text{in/out}$, it will not enclose precisely a half of luminosity of the spiral, and the specific fraction of light enclosed will depend on $n^\text{in/out}$. Therefore, we use the FWHM as a measure of width of the spiral arm, which is simply a sum of $w_\text{in}$ and $w_\text{out}$. These values are measured from the ridge-line of the spiral arm to locations of half-maximum intensity along the radius.

\par

\begin{figure}
\centering
\includegraphics[width=0.95\columnwidth, angle=0]{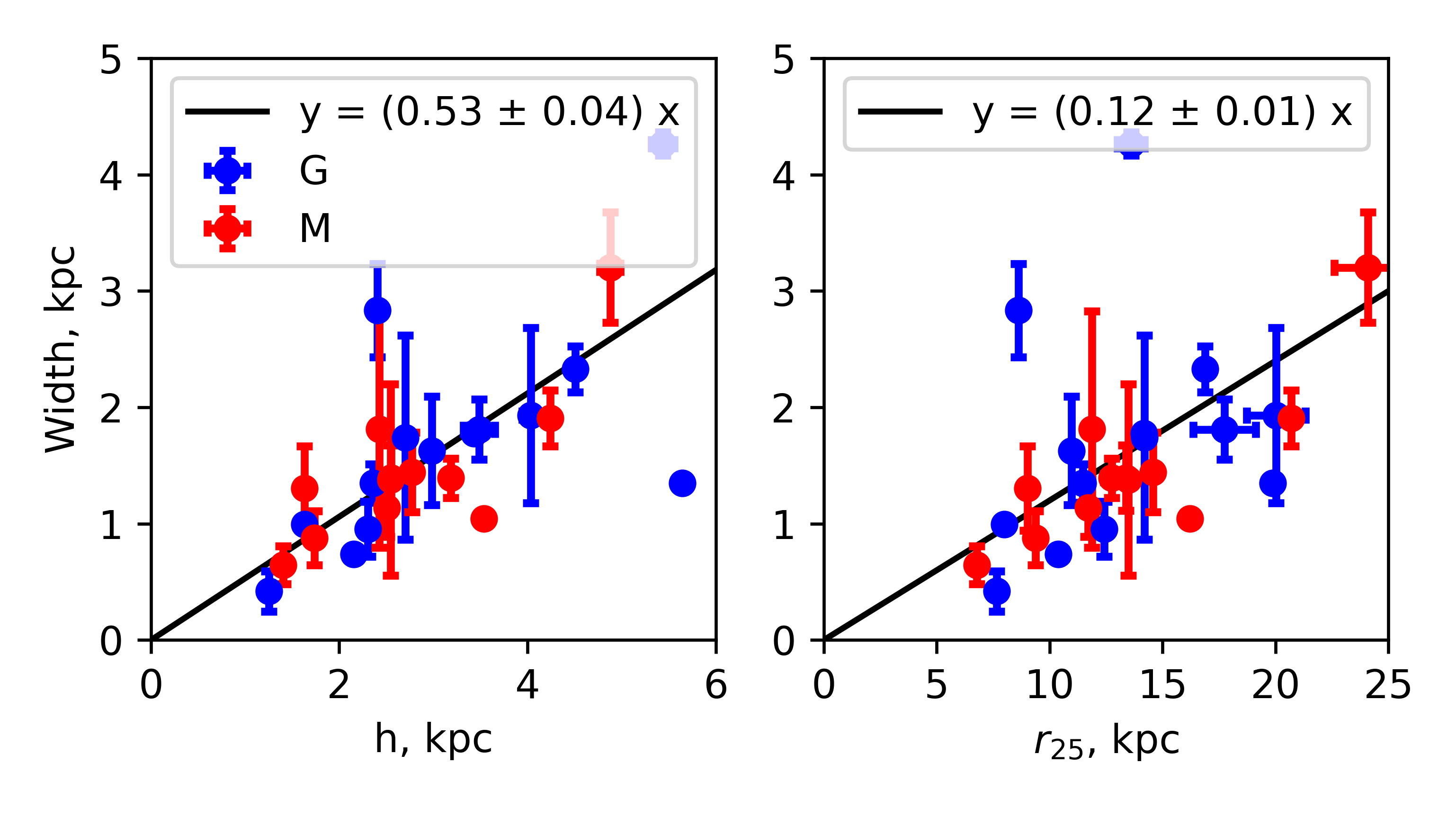}
\caption{Relation between the width of spiral arms and disc scale length $h$ (left) and radius $r_{25}$ (right)}
\label{fig:h_width}
\end{figure}

Using the result of our fitting, we find a linear relation between the width of spiral arms $w$ and disc scale length $h$ and disc optical radius $r_{25}$. As seen in Fig.~\ref{fig:h_width}, $w = (0.53 \pm 0.04)~h$ and $w = (0.12 \pm 0.01)~r_{25}$, on average. For the $r$ band,~\citet{Savchenko2020} found $w = 0.16~r_{25}$, although the authors used a different method to measure the width which appears to be larger for the same distribution of light. Moreover, the width of galaxy spiral arms is not constant with wavelength and Marchuk et al. (in prep.) found for M\,51 that the width of spiral arms is higher in the $r$ band than at 3.6 $\upmu$m. 

\begin{figure}
\centering
\includegraphics[width=0.95\columnwidth, angle=0]{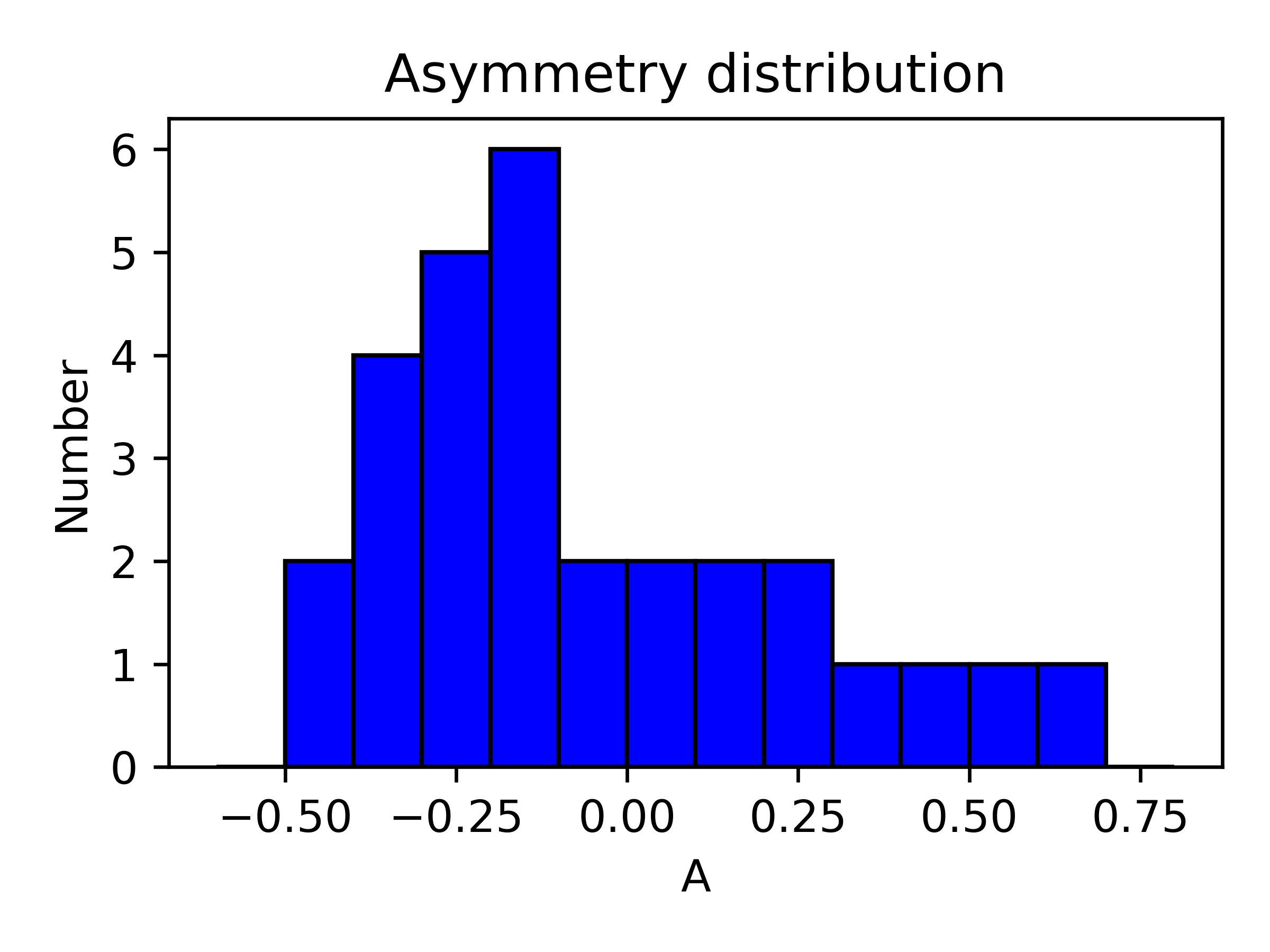}
\caption{Distribution of the asymmetry of spiral arms' perpendicular profile. If outer part extends further from the spiral arm ridge, the value is positive.}
\label{fig:spiral_asymm}
\end{figure}

The asymmetry of spiral arms can also be measured. We express the asymmetry in terms of the relative difference between the inner and outer widths $A = \frac{w_\text{out} - w_\text{in}}{w}$, so that $A$ ranges from $-$1 to 1 in most extreme cases possible and equals 0 when the arm is symmetric. The distribution of $A$ is shown in Fig.~\ref{fig:spiral_asymm}, and the mean value of $A$ for galaxies in our sample is close to zero but negative, namely $-0.05$. This means that there is a weak systematic asymmetry, with the inner part of spiral arms being more extended than the outer part, with respect to the spiral arm ridge. Interestingly, the distribution of $A$ looks highly asymmetric and about 2/3 of galaxies have negative $A$ but the range of possible values of $A$ is more extended to positive values than to negative (there are no galaxies with $A < -0.5$ but $A > 0.6$ is possible). \citet{Savchenko2020} found a positive average asymmetry indicating that the outer part of spiral arms is usually more extended than the inner one. They measure asymmetry as $A' = = \frac{w_\text{out} - w_\text{in}}{w_\text{out}}$ which we consider less convenient and our values cannot be compared directly with theirs but, nevertheless, we obtain the qualitatively opposite conclusion which should not change if we adopt their definition of asymmetry. However,~\citet{Savchenko2020} used $r$-band images for their analysis, so our and their results regarding asymmetry may differ significantly, because different photometric bands highlight different populations of stars in galaxies, which may be distributed differently inside the spiral arm.

\begin{figure}
\centering
\includegraphics[width=0.95\columnwidth, angle=0]{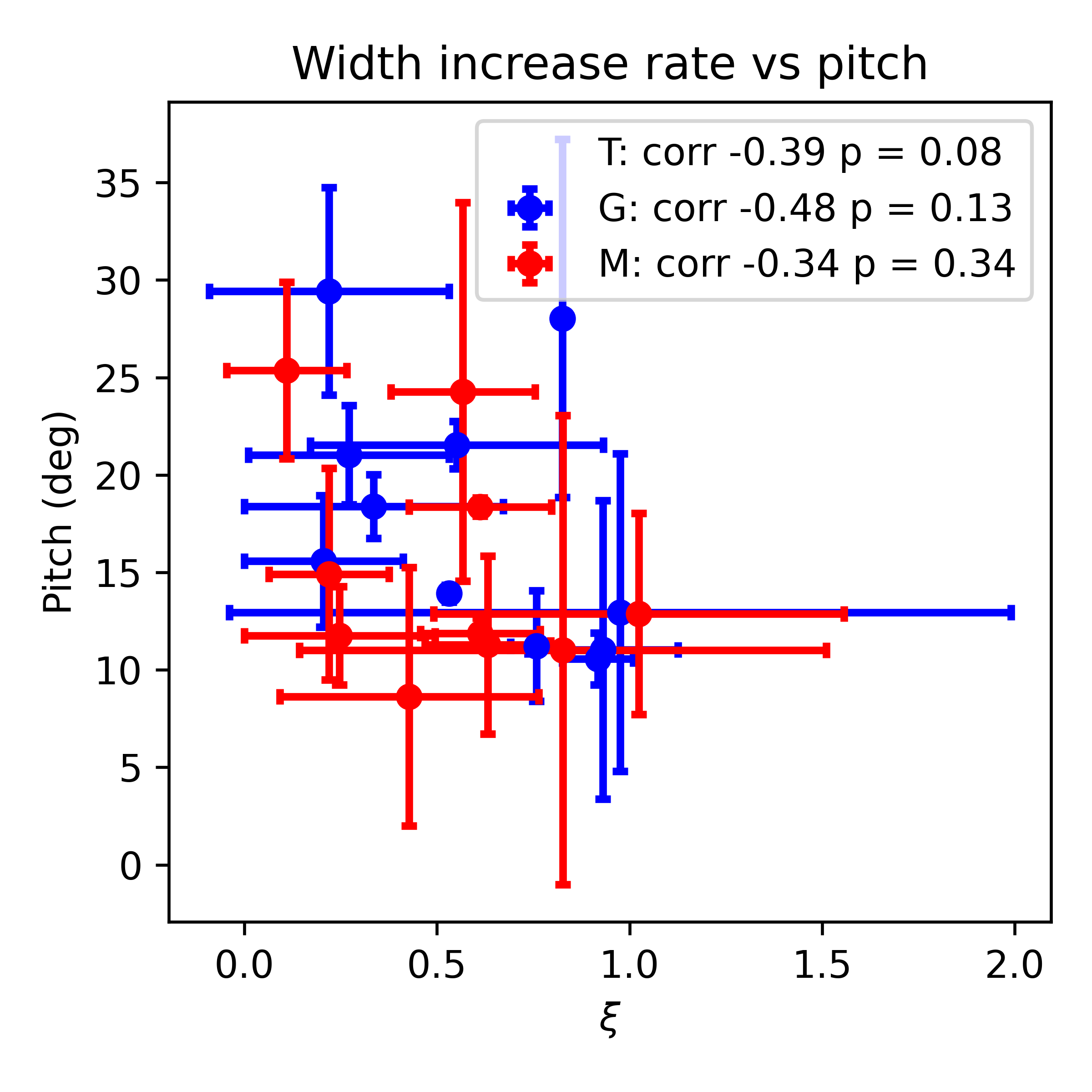}
\caption{Diagram showing the pitch angle versus the spiral arm width increase rate $\xi$}
\label{fig:xi-pitch}
\end{figure}

We also examine the $\xi$ value which determines the spiral arms width increase rate. This value is restricted to non-negative values. In Fig.~\ref{fig:xi-pitch}, we show pitch angles plotted versus $\xi$ for $0 < \xi < 1.1$. A very subtle but clear trend can be seen that width increases more rapidly in more tightly wound arms. We should note that parameter $\xi$ itself has a rather inconvenient dimension of $\text{arcsec}^{2/3} / \text{rad}$ (see eq.~\ref{eq:I_bot}), though this parameter allows one to obtain a good approximation of the spiral arm's shape. Most individual spiral arms have a non-zero positive $\xi$, indicating that their width is not constant and increases to periphery. On average, the arm width at the beginning is 74\% of the arm width at the end. Hydrodynamical simulations in~\citet{Forgan2018} do not reproduce any noticeable width alteration with radius, whereas~\citet{Savchenko2020} measurements show that width in most cases increases with radius, which is in agreement with our results. \citet{Honig2015} measured the arm properties via the distribution of HII regions and also found that the arm width increases outwards. Interestingly, they found a reversal of this trend in the periphery of the arm, which cannot be reproduced with our model.

We also examine the scatter of individual spiral arms' width in each galaxy, measured as a ratio between the standard deviation of spiral arm width $\sigma_\text{w}$ and their mean value $\langle w \rangle$, see Table~\ref{tab:spiral_parameters}. We find that $\sigma_\text{w} / \langle w \rangle$ is 23\%, on average, indicating that width variation in a single galaxy is usually not very high. We notice that the average $\sigma_\text{w} / \langle w \rangle$ is only 10\% for two-armed galaxies (with $\sigma_\text{w}$ being simply a half of width difference between two arms in this case), and 37\% for others. The lower $\sigma_\text{w} / \langle w \rangle$ for two-armed galaxies agrees with the fact that grand-design galaxies are known to have more symmetric spiral structure. 

\subsection{Connection with bar parameters}
Various studies point to a connection between spiral arms and bars in galaxies \citep{Athanassoula10,Minchev12}. Most of the galaxies in our sample are barred, and different observed or predicted relations can be verified.

\par

\begin{figure*}
\centering
\includegraphics[width=1.95\columnwidth, angle=0]{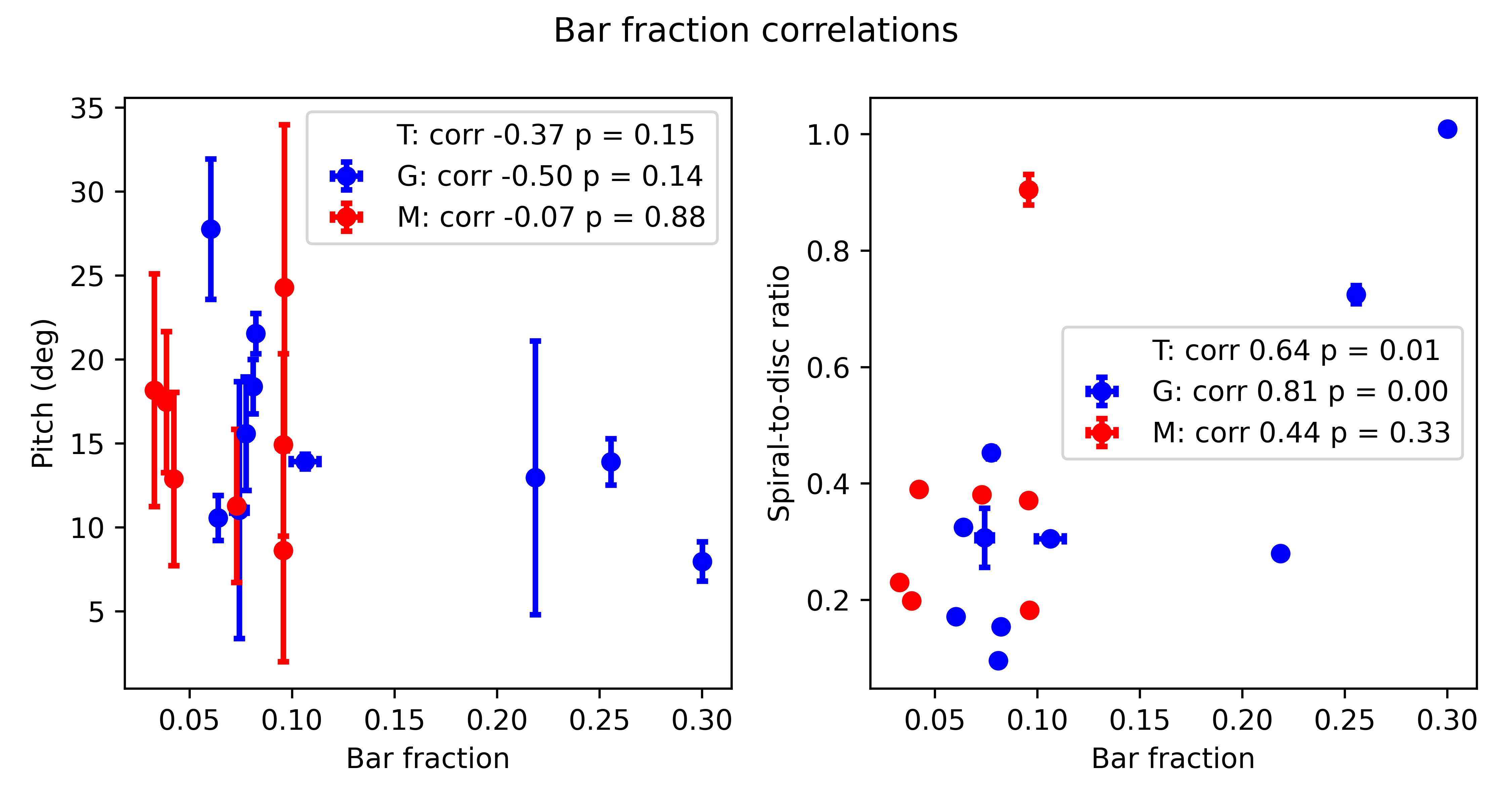}
\caption{Correlations of bar fraction and spiral properties (only barred galaxies are included).}
\label{fig:bar_corrs}
\end{figure*}

We find that pitch angle weakly decreases and spiral-to-disc ratio ($S/D$) increases with increasing bar-to-total ratio ($\text{Bar}/T$), see Fig.~\ref{fig:bar_corrs}. Interestingly, if $S/T$ is plotted against $\text{Bar}/T$, the correlation will be slightly negative, in contrast, at first sight, to the positive $\text{Bar}/T$-$S/D$ correlation. We can explain it by the fact that strongly-barred galaxies obviously have a smaller fraction of the total flux contained in the other components except the bar and spirals, and $S/D$ for strongly-barred galaxies will be significantly higher than $S/T$. Moreover, three galaxies with strongest bars in our sample, namely NGC~986, NGC~4314, and NGC~4548, have a complex structure consisting of other components besides disc, bulge, and bar, making $S/T$ for these galaxies even smaller compared to $S/D$.

Concerning the relation between pitch angle and bar-to-total ratio,~\citet{Diaz-Garcia2019} did not find any correlation between these quantities, whereas the results from~\citet{Font2019} are consistent with ours. Both~\citet{Bittner2017} and mentioned~\citet{Diaz-Garcia2019} found a positive correlation between spiral contrast and bar contrast, which agrees with our result. Note that only grand-design galaxies in our sample host strong bars, whereas multi-armed galaxies have $\text{Bar}/T < 0.1$ and more often have no bar. This is consistent with~\citet{Hart2017} and again points to a strong connection between the bars and spirals.


\subsection{The light distribution in spirals along the radius}

\begin{figure*}
\centering
\includegraphics[width=1.95\columnwidth, angle=0]{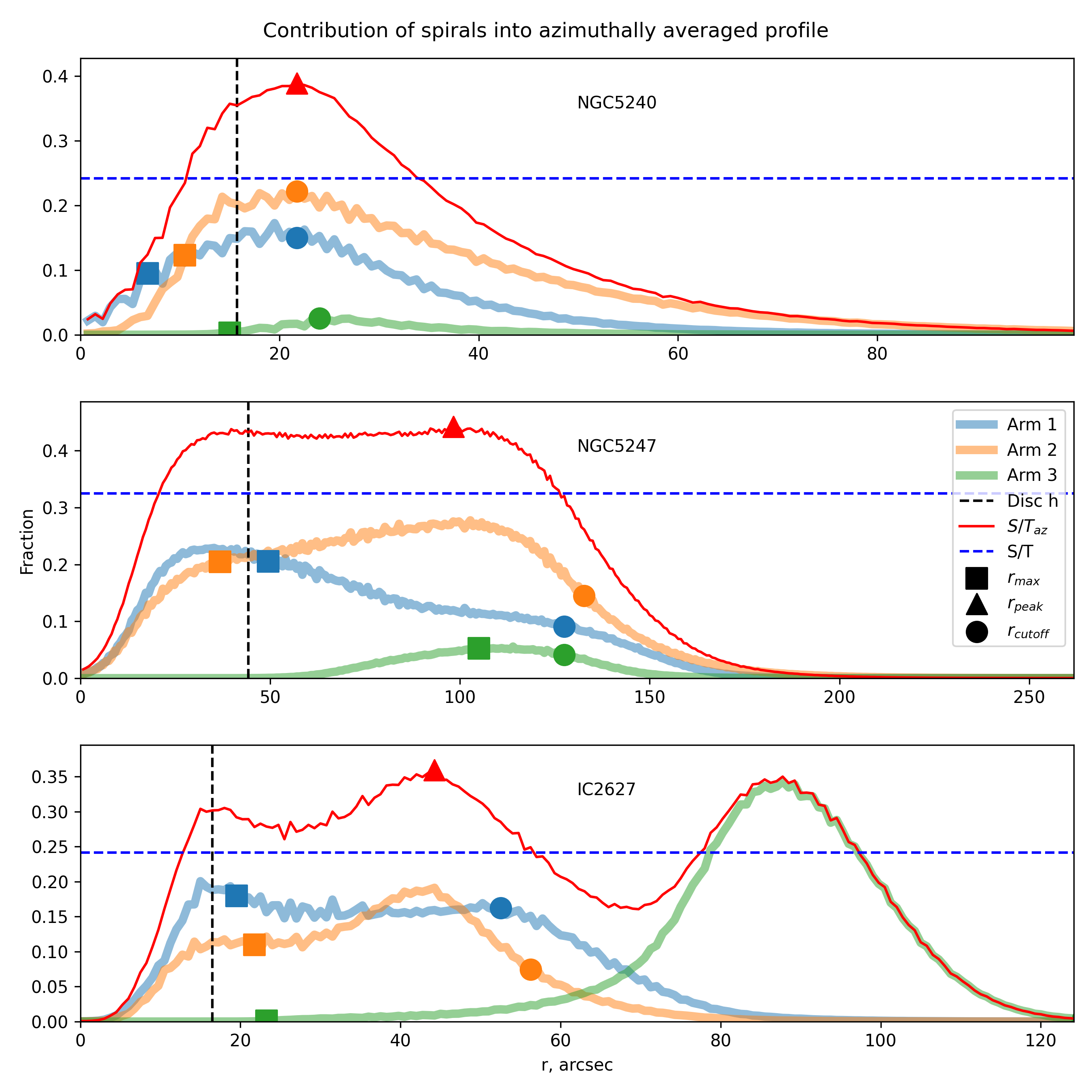}
\caption{Examples of azimuthally averaged profile of spiral arms contribution, expressed as a $S/T_\text{az}(r)$ function. Each galaxy is an example of one of three main types of profiles. The function $S/T_\text{az}(r)$ for NGC~5240 has a single peak, for NGC~5247 it has a ``plateau'', and for IC~2627 multiple peaks are seen. Similar images for other galaxies can be found in Appendix.}
\label{fig:az_examples}
\end{figure*}

In this subsection, we analyze the contribution of the spiral arms to azimuthally-averaged surface brightness profiles of the sample galaxies. For this purpose, we use a function $S/T_\text{az}(r)$, which represents the fraction of light provided by the spiral arm model at radius $r$. In Fig.~\ref{fig:az_examples}, we give a few examples of $S/T_\text{az}(r)$ for galaxies in our sample.

We find that in almost all cases the function $S/T_\text{az}(r)$ reaches its maximum value at a moderate $r$, decreasing to zero in the center and in the periphery of the galaxy. The only exception is NGC~2460, where $S/T_\text{az}(r)$ continues to increase far beyond the disc and reaches almost unity. We can conclude that in the vast majority of galaxies, spiral arms are truncated at smaller distances than the disc is. For NGC~2460, the opposite behaviour can be explained by the presence of IC~2209 in its neighborhood, and their possible interaction might have formed the extended tidal arms in the NGC\,2460. The drop of $S/T_\text{az}(r)$ to zero near the center in all cases is easily explained by the fact that spiral arms are not observed in the very center of galaxies and bulge contributes the most in the center.

\par

\begin{figure}
\centering
\includegraphics[width=0.95\columnwidth, angle=0]{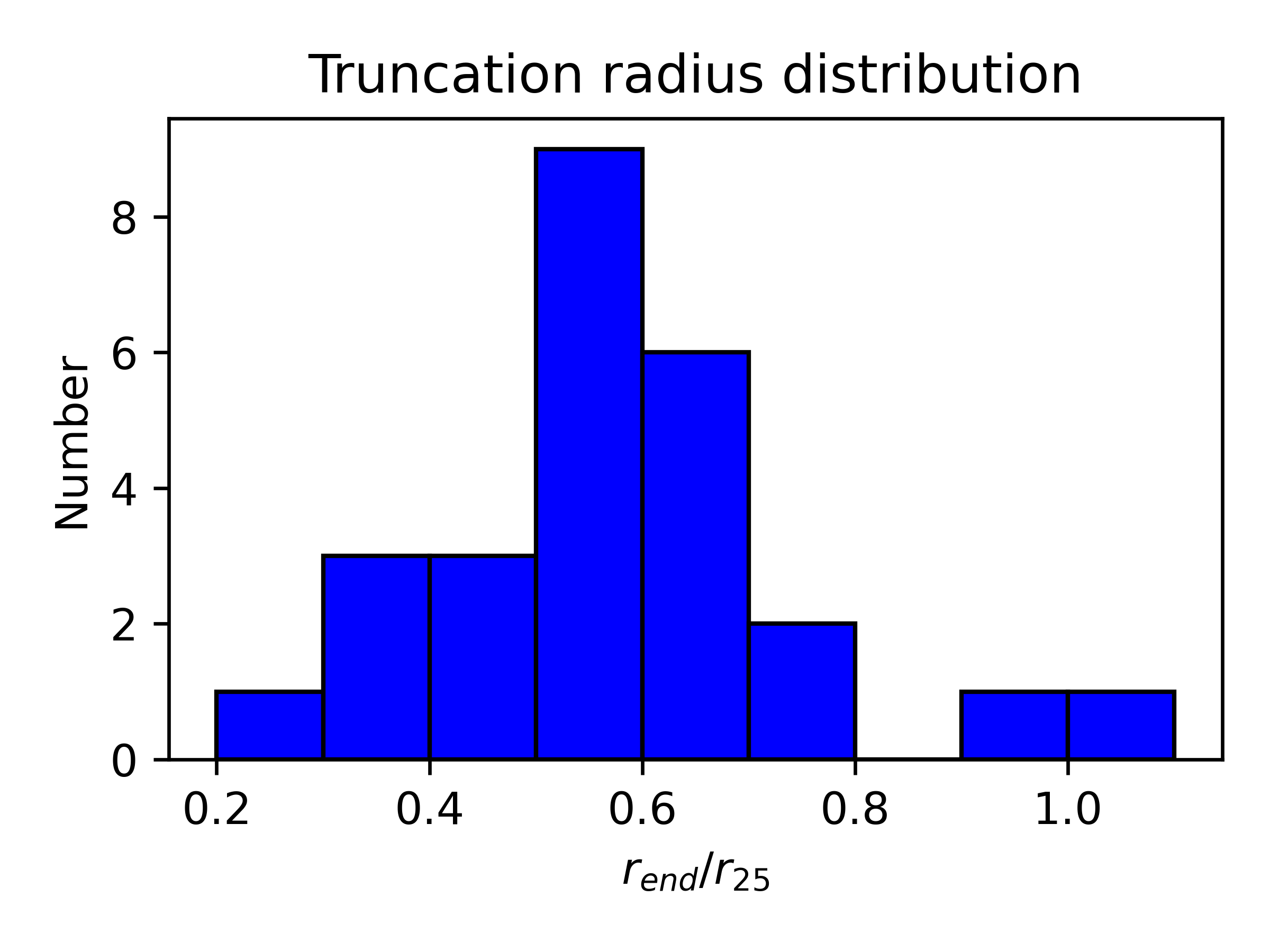}
\caption{Distribution of the truncation radii $r_\text{end}$ relative to the optical radii of galaxies $r_{25}$.}
\label{fig:r_truncation}
\end{figure}

There are many galaxies with average $h_\text{s} / h$ larger than unity (see Table.~\ref{tab:spiral_parameters}), which means that spiral arms may have higher exponential scale than the disc and, thus, fade slower with radius in some part of disc. However, almost all of the galaxies have their arms truncated at some radius, which is apparently caused by the termination of star formation when gas density falls below the critical value~\citep{Kregel2004}. For our sample, the average truncation radius of spiral arms in a galaxy $r_\text{end}$ usually lies between 0.5 and 0.7 of its optical radius $r_{25}$ (Fig.~\ref{fig:r_truncation}) but the full range of possible values of $r_\text{end} / r_{25}$ is large. Such a truncation is thought to be one of the reasons for the appearance of disc downbending profiles, i.e. transitions to smaller disc radial scale length at certain radius~\citep{Bittner2017}. Various studies find that from 20\%~\citep{Hunter2006} to 60\%~\citep{Pohlen2006} of disc galaxies exhibit down-bending (Type II) profiles. At the same time, our sample contains only three galaxies, namely NGC~2460, NGC~4548, and NGC~5247, whose discs cannot be modelled with a single exponential function when spiral arms are included in the model, and, instead, are fitted with a broken exponential. Only the latter two have genuine down-bending profiles apparently not connected with spiral arms, which is less than 10\% of our sample. 
Since we expect a larger fraction of such profiles~\cite{Hunter2006,Pohlen2006}, we cautiously conclude that the truncation of spiral arms may be connected with down-bending of disc profiles (see Mosenkov et al., in prep.). 

We also show that the behaviour of $S/T_\text{az}(r)$ near its maximum also varies between galaxies, see Fig.~\ref{fig:az_examples} again. We distinguish three main shapes of this function. The first type is one well-localized maximum, as in NGC~5240. The second type is a ``plateau'', i.e. an extended zone where $S/T_\text{az}(r)$ is nearly constant, as in NGC~5247. The last type is characterized by multiple local maxima, usually produced by separate spiral arms, as in IC~2627.

\par

The analysis of the Fourier modes' amplitude relative to axisymmetric components in~\citet{Kendall2011} closely resembles our analysis of azimuthally-averaged profiles. The amplitude of the Fourier modes in their analysis varies with radius and usually decreases near the center, in agreement with our results. However, they traced amplitudes only out to the optical radius or even to a fraction of it, so we cannot compare our and their results regarding the decrease of $S/T_\text{az}(r)$ in the periphery. We have only one common galaxy with their study, namely NGC~5194 (M\,51), and they traced it up to about 275 arcsec from the center (see fig.~45 in~\citealt{Kendall2011}). In this range, our results are roughly consistent with theirs.

\par

\begin{figure*}
\centering
\includegraphics[width=1.95\columnwidth, angle=0]{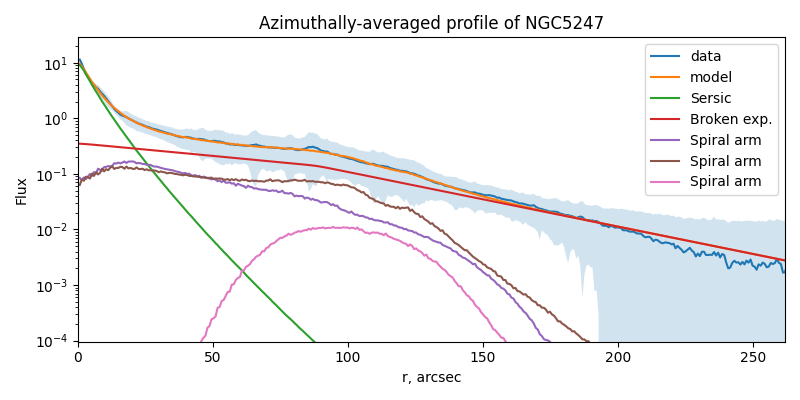}
\caption{Azimuthally averaged profile of NGC~5247. At $r=100\ldots150$~arcsec, one can see that the spiral arms are truncated and the surface brightness of the whole galaxy becomes disc-dominated. At smaller $r$, the spirals make the overall profile brighter than the pure exponential, which is manifested as a ``bump'' on the profile.}
\label{fig:NGC5247-az-avg}
\end{figure*}

We will now discuss azimuthally averaged profiles of galaxies as a whole. In Fig.~\ref{fig:NGC5247-az-avg}, one can see a profile of NGC~5247, shown as an example. In the periphery of the galaxy, where the truncation of the spiral arms occurs, only the disc component contributes to the surface brightness because the profile is purely exponential. Let us now consider the radius $r_\text{peak}$ where $S/T_\text{az}(r)$ reaches its maximum. We can say that only the spiral arms and the disc are contributing to the profile at $r_\text{peak}$ because the contribution of the central components, such as the bulge, is insignificant in the region where the spiral arms are most prominent. We can define $T/D_\text{az}(r)$ as a function representing the ratio of the azimuthally-averaged surface brightness of the galaxy and the same but for the disc model only. It is known that spiral arms create a ``bump'' over a pure exponential surface brightness profile of the disc, for example, see~\citet{Casasola2017}. The prominence of this bump can be expressed as $T/D_\text{az}(r_\text{peak})$, since we have shown that only disc and spiral arms are contributing to the profile at $r_\text{peak}$, and the contribution of the spiral arms is the highest there. It is thus natural to assume that the prominence of the bump is related to the spiral-to-total luminosity ratio.

\begin{figure}
\centering
\includegraphics[width=0.95\columnwidth, angle=0]{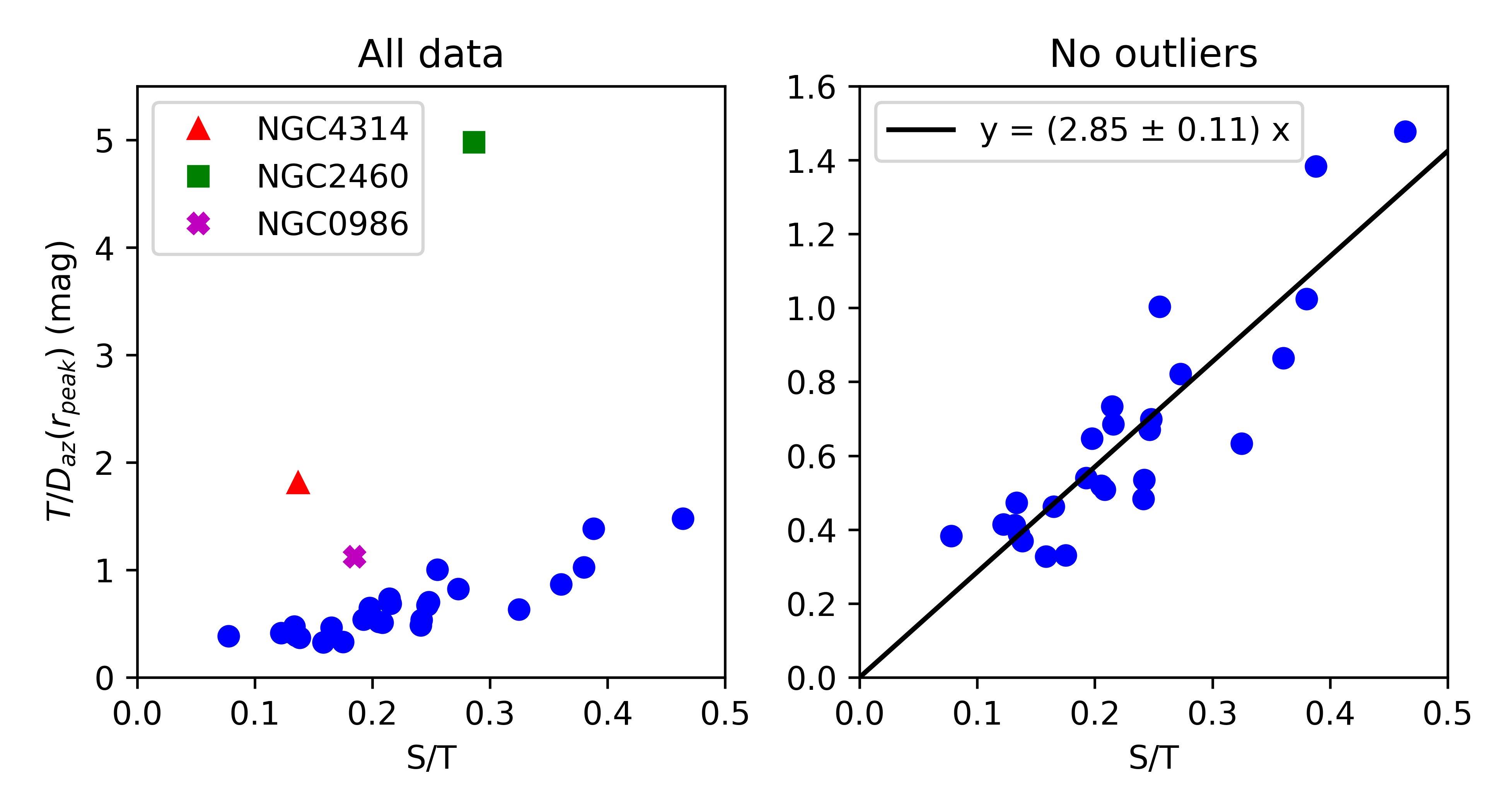}
\caption{Relation between the $T/D_\text{az}(r_\text{peak})$ and $S/T$. Values for all galaxies in the sample are shown on the left, outliers are marked with non-circular symbols. The same relation with outliers excluded and linear approximation shown is on the right.}
\label{fig:T-D}
\end{figure}

In Fig.~\ref{fig:T-D}, we show the relation between $T/D_\text{az}(r_\text{peak})$, expressed in magnitudes, and $S/T$. We see three outliers from this relation, of which the aformentioned NGC~2460 stands out the most. It can be excluded from the relation because its spiral arms extend farther than the disc and so their contribution in the periphery is much higher than the disc's contribution. The other two outliers are NGC~986 and NGC~4314 which have luminous central components, bars, and rings and, therefore, their disc-to-total luminosity ratios are low. At the same time, spiral arms in these galaxies form outer pseudorings, and these contribute mostly at large radii, where the discs are faint. If we exclude these outliers, we will obtain the relation $T/D_\text{az}(r_\text{peak}) (\text{mag}) = (2.85 \pm 0.11)~S/T$. Assuming that $S/T$ is usually between 0.1 and 0.25, the typical bump on the azimuthally-averaged profile atop the pure exponential is 0.3--0.7 mag, but we see that for galaxies with the most prominent spiral structure it exceeds 1.4 mag.

\begin{figure}
\centering
\includegraphics[width=0.95\columnwidth, angle=0]{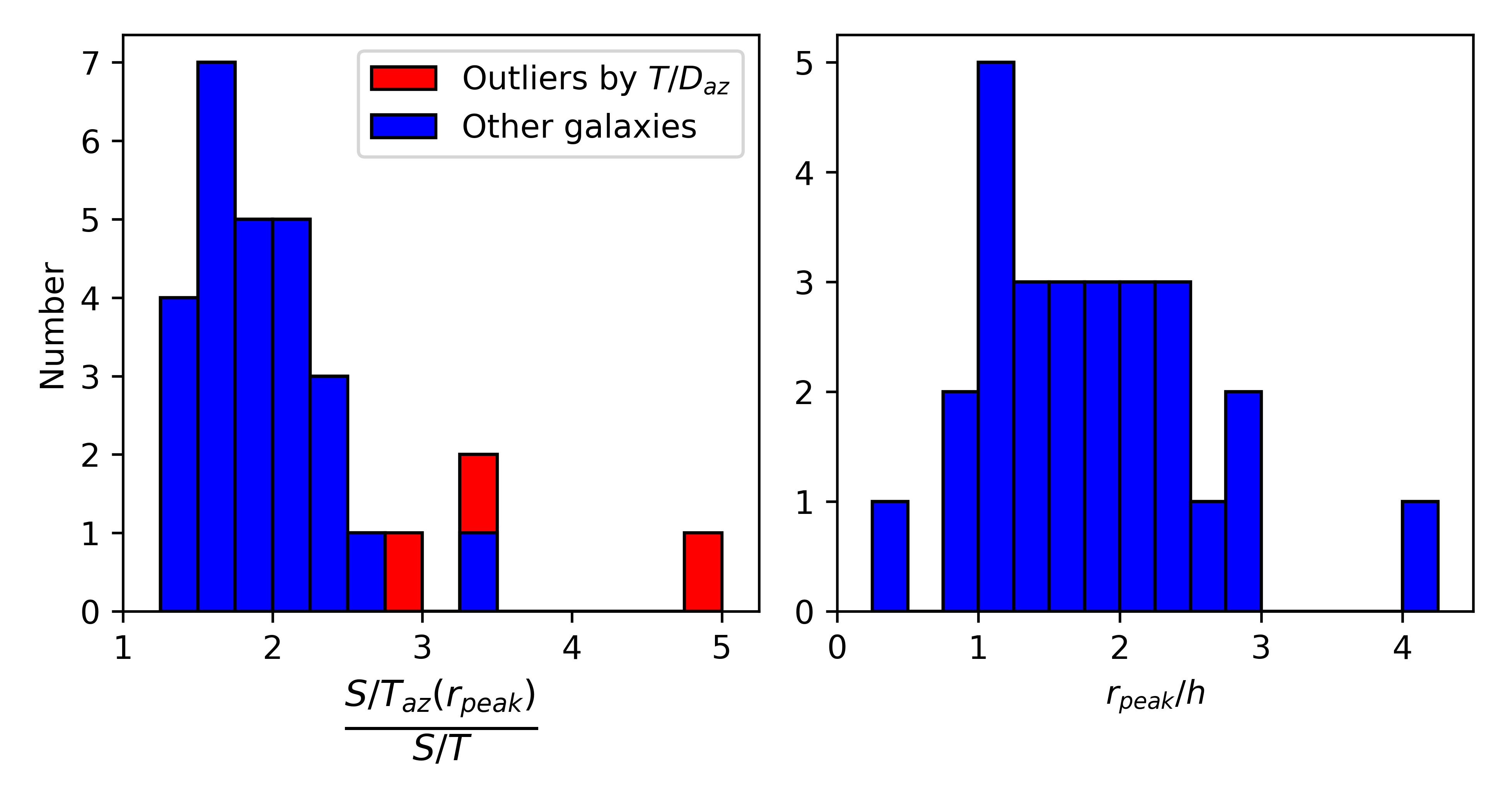}
\caption{On the left, the distribution of the ratio of $T/D_\text{az}(r_\text{peak})$ to $S/T$ is shown, with the outliers from Fig.~\ref{fig:T-D} marked with red. On the right, the distribution of the ratio of the $r_\text{peak}$ to disc exponential scale is shown.}
\label{fig:peak_T-D}
\end{figure}

In Fig.~\ref{fig:peak_T-D}, the left-hand side, we show a distribution of $S/T_\text{az}(r_\text{peak}) / (S/T)$, i.e. the ratio between the highest contribution of the spiral arms to the azimuthally-averaged profile and the overall spiral-to-total ratio, which lies in most cases between 1.5 and 2.5. In Fig.~\ref{fig:peak_T-D}, the right-hand side, we show a distribution for $r_\text{peak} / h$. We find that the contribution of the spiral arms to the azimuthally-averaged profile is usually the highest at a distance of 1--2 disc radial scale lengths from the center. 

\section{Discussion}
\label{sec:discussion}

We have found that the parameters of galactic structural components, derived by means of photometric decomposition, are prone to systematic errors when spiral arms are not accounted for. This was shown in~\cite{Sonnenfeld2022} by the analysis of some model galaxies, while here we demonstrate the same result for a sample of real galaxies. The inclusion of spiral arms in the model is especially important for the disc brightness. We have also found that the presence of spiral arms produces a ``bump'' on the galaxy azimuthally-averaged surface brightness profile. This means that if one measures the parameters of a spiral galaxy without treating the spiral arms properly, the estimates may be erroneous, even if one uses other methods than decomposition, e.g. azimuthally-averaged profile analysis. This may have far-reaching consequences. In particular, various known scaling relations for spiral galaxies or their components~\citep{D'Onofrio2021} may need to be reconsidered to take this effect into account. Most obviously, the known distribution of the central disc surface brightness $I_0$~\citep{O'Neil2000} probably should be adjusted, because its value is commonly determined by extrapolation of the disc brightness profile to the center. Assuming that the height of the mentioned ``bump'' is usually 0.3--0.7~mag at 3.6$\upmu$m, the required correction to the measured $\mu_0$ should be of the same order of magnitude.

\par
Apart from introducing biases in the scaling relations, neglecting spiral arms in the decomposition leads to significant errors at the level of individual galaxies. 
 Therefore, a proper treatment of spiral arms may reduce the uncertainty of these parameters and minimize the scatter in physically significant scaling relations. For example, a photometric distinction between classical bulges ($n$ > 2) and pseudobulges ($n$ < 2) is well-known (see e.g. \citealt{Gadotti2009}), and different scaling relations demonstrate the difference of these two types of bulges~\citep{Gadotti2009,Fisher2010}. Since estimation of the S\'ersic index $n$ in decomposition can be affected by the presence and brightness of spiral arms in a galaxy, it is possible that some of those bulge scaling relations can be influenced to some degree as well.

\par

Photometric decomposition is the most straightforward way to distinguish the light from spiral arms and main galaxy components, if a suitable model of spirals is used. At the same time, decomposition with spiral arms allows one to measure a large set of parameters of spiral arms. The main problem is that such method is time-consuming and not easy to implement, which limits the possible sample size. Nevertheless, our results show various connections between some parameters of spiral arms and other components, implying that it is possible to establish some scaling relations for spiral arms and define a small number of key parameters of spiral structure related to other structural parameters and general properties of galaxies. In this paper, we have determined some of these key parameters. For instance, the spiral-to-total luminosity ratio seems to be important for understanding the origin of different types of spiral structure.


\par

Previously, decomposition was not used widely for the determination of spiral arms parameters. Therefore, it is important to examine the general consistency of our results with other methods used in the literature. The good starting point is the study by~\citet{Diaz-Garcia2019}, since our subsample of galaxies was selected from their much larger sample, and we utilize the same S$^4$G data. They measured pitch angles of spiral arms using a Fourier technique. A quick comparison with our results shows that our average pitch angles are roughly consistent with theirs for individual galaxies. Moreover, the agreement between our results and~\citet{Diaz-Garcia2019} is not worse than the agreement between their results and other studies employing the same Fourier technique (see, for example, fig.~5 in their paper). We also compare our results with \citet{Savchenko2020} since they measured various parameters of spiral arms using a very direct method of making and analyzing perpendicular slices of spiral arms. Although they used a different set of galaxies and different  wavelengths ($gri$), our conclusions concerning the width of spiral arms and their contribution to the total luminosity are remarkably consistent with theirs. This proves that our complex modeling of galaxy structure with spiral arms is a credible approach to determine the parameters of spiral structure.

\par

Decomposition, if done for a multiband set of galaxy images, can also be important for separately studying resolved spectral energy distributions (SEDs) of different galaxy subsystems, such as a disc and a bulge. This makes it possible to reconstruct the star formation rate and assembly history of these components. This approach has been successfully applied to the lenticular galaxy NGC~3115 by \citet{2021MNRAS.504.2146B} and more recently for M~81 in \cite{2023ApJS..267...26G}. In the latter case, it is easy to see in the residuals map (Fig.~17 and Fig.~19) that the enormous spiral arms of M~81 were not included in their model. The spiral arms should usually have a similar stellar history as the underlying stellar disc. However, our results for 3.6~$\mu$m in this paper and for a multiwavelength dataset for M~51 in Marchuk et al. (in prep) also demonstrate that the SED of the bulge may be affected, if spiral arms are not properly accounted for in modeling. Therefore, in future studies, these considerations should be taken in account when spiral arms have a significant contribution to the galaxy SED.

\par

Our analysis has shown that spiral arm model, which is adopted in our work, is suitable for fitting spiral arms. However, it has some drawbacks which were revealed when applying it in practice. Some of these issues are mostly technical and their correction will not alter the form of the functions that defines our model. At first, this concerns the exact set of values which are used as input parameters of functions in {\small{IMFIT}} input files. For example, instead of the pair of $\varphi_\text{cutoff}$ and $\varphi_\text{end}$, we can redefine the function to use the equivalent pair of $\varphi_\text{cutoff}$ and $\varphi_\text{end} - \varphi_\text{cutoff}$, where the latter has physical meaning of length of the cutoff region. However, the first pair of values is more convenient to use because the position of $\varphi_\text{cutoff}$ and the length of the cutoff region has some degeneracy when both used as free parameters.

\par
Even though our model has a large number of free parameters and is flexible enough to properly account for intricate details of spiral arms, such as a variable pitch angle and widening, the application of our model to real galaxies revealed that there is still room for improvement of the model.   
For example, in our model, the half-width can only increase on the inner and outer sides of the arm with the same rate. However, it would be more accurate to consider independent rates of the width's change on both sides of the arm. This can be justified by the following physical reasons. Because the spiral arms in a galaxy rotate faster than the disc outside the corotation radius and slower inside of it, the matter enters a spiral arm from different sides in the central and peripheral parts of the galaxy. When material enters the arm, star formation occurs. The formed young stellar population then leaves the arm on the opposite side of the arm (see e.g. \citealt{2023MNRAS.524...18M}), and the ``trail'' of young stars makes one side more extended then another. For a trailing spiral arm, in the central part of galaxy the outward side will be more extended, and in the periphery the opposite will be true~\citep{Marchuk2023}. Therefore, a half-width of the spiral arm  can indeed change for the inward and outward sides of the arm independently. In future, we plan to improve our model and use a greater number of bands. It will allow us to locate corotation radii in galaxies~\citep{Marchuk2023}, which is of great importance for studying the dynamics of galaxies. 
\section{Conclusions}
\label{sec:conclusions}

Galaxies with a prominent spiral pattern are ubiquitous in the local Universe, but spiral arms are rarely taken into account in the photometric decomposition of galaxies. In this paper, we have attempted to remedy this lack and draw the following conclusions:

\par
\begin{enumerate}
\item  We have applied a new photometric model of spiral arms, which was developed in the accompanied paper Marchuk et al (in prep.) and describes the 2D distribution of surface brightness independently for each spiral arm. Our function has an advantage that all its parameters have a clear physical meaning. It can accurately reproduce real spiral arms with a different geometry, including those with variable widths and pitch angles, and with a different distribution of light both along and across the arm. Using this function, we have performed decomposition of 29 spiral galaxies which is enough to conduct a statistical analysis of our results.

\item We have compared ``classical'' models consisting only of commonly used components (disc, bulge, bar, etc) and models with spiral arms for each galaxy. We measured how neglecting the spiral arms affects the estimation of parameters in our sample. After including the spiral arms in the model, surface brightness of the disc decreases by 0.5 mag at average. We found that parameters of the bulges and bars, as well as the disc exponential scale, are also change significantly when including spiral arms, however in most cases these changes are different for different galaxies. 

\item We have confirmed that spiral arms can contribute significantly to galaxy luminosity. Their contribution is usually 10\%--25\% but, in some galaxies, it may exceed 45\% of the total luminosity in the 3.6 $\upmu$m band. We found that spiral arms contribution is higher for galaxies with a higher bulge-to-total ratio and which host more luminous discs, as well as for galaxies of intermediate Hubble types (see Figs.~\ref{fig:T_fracs} and~\ref{fig:ST_corrs}). Some of these results have been reported by~\citet{Savchenko2020}, but in this paper we have obtained them using the accurate photometric decomposition for the first time.

\item Our method has allowed us to measure pitch angles for each spiral arm independently and also trace the variability of the pitch angle along the spiral arm. We found that the pitch angle variation in a single spiral arm is $8^\circ$, on average. Therefore, we conclude that spiral structure in galaxies cannot be generally characterized by a single pitch angle, or even by average pitch angles of individual arms. We confirm the weak anticorrelation between pitch angle and bulge fraction, which was found in some previous studies, but using our new method. We confirm that the pitch angle variation is the smallest in galaxies with the most prominent spirals, again in consistency with~\citet{Savchenko2020}.

\item We have measured the widths of the spiral arms and find that the width, expressed in terms of FWHM, in on average equal to 53\% of the disc scale length or 12\% of the disc optical radius (see Fig.~\ref{fig:h_width}). We found a weak systematic asymmetry of the perpendicular profiles of spiral arms, with inner parts being little more extended than outer. We measured the rate of width increase along the spiral arms and found that the width at the beginning of arms is 73\% of the width at the end of arms, on average.

\item We have inspected the connection between the parameters of the bar and the spiral arms, and found that pitch angle weakly decreases and spiral-to-disc ratio increases with an increase of the bar fraction (see Fig.~\ref{fig:bar_corrs}). We also confirm that strongly barred galaxies tend to have grand-design spiral structure.

\item We have analyzed the contribution of the spiral arms to the azimuthally-averaged galaxy profiles. In almost all cases, the contribution of spiral structure reaches its highest value at moderate radii, usually at 1--2 disc scale lengths from the center, decreasing to zero near the center and in the periphery of the galaxy. We found a relationship between the spiral-to-total luminosity ratio and the size of the ``bump'' on the azimuthally-averaged profile associated with the presence of the spirals. We found its typical value to be 0.3--0.7~mag.
\end{enumerate}

\par
Overall, we conclude that various estimates of the galaxy parameters in the literature, which do not account properly for the presence of spiral arms, are likely to be biased. This may have far-reaching consequences for some scaling relations of spiral galaxies and, thus, such relations need to be updated using accurate photometric decomposition with spiral arms included.

\section*{Acknowledgements}
We acknowledge financial support from the Russian Science Foundation, grant no. 22–22–00483.\\
This work is based on observations made with the Spitzer Space Telescope, which was operated by the Jet Propulsion Laboratory, California Institute of Technology under a contract with NASA.\\
This research has made use of the NASA/IPAC Extragalactic Database (NED), which is operated by the Jet Propulsion Laboratory, California Institute of Technology, under contract with the National Aeronautics and Space Administration.\\
We acknowledge the usage of the HyperLeda database\footnote{\url{http://leda.univ-lyon1.fr}}.\\

\section*{Data availability}
The data underlying this article will be shared on reasonable request to the corresponding author.

\bibliographystyle{mnras}
\bibliography{art}

\begin{thebibliography}{}
\makeatletter
\relax
\def\mn@urlcharsother{\let\do\@makeother \do\$\do\&\do\#\do\^\do\_\do\%\do\~}
\def\mn@doi{\begingroup\mn@urlcharsother \@ifnextchar [ {\mn@doi@}
  {\mn@doi@[]}}
\def\mn@doi@[#1]#2{\def\@tempa{#1}\ifx\@tempa\@empty \href
  {http://dx.doi.org/#2} {doi:#2}\else \href {http://dx.doi.org/#2} {#1}\fi
  \endgroup}
\def\mn@eprint#1#2{\mn@eprint@#1:#2::\@nil}
\def\mn@eprint@arXiv#1{\href {http://arxiv.org/abs/#1} {{\tt arXiv:#1}}}
\def\mn@eprint@dblp#1{\href {http://dblp.uni-trier.de/rec/bibtex/#1.xml}
  {dblp:#1}}
\def\mn@eprint@#1:#2:#3:#4\@nil{\def\@tempa {#1}\def\@tempb {#2}\def\@tempc
  {#3}\ifx \@tempc \@empty \let \@tempc \@tempb \let \@tempb \@tempa \fi \ifx
  \@tempb \@empty \def\@tempb {arXiv}\fi \@ifundefined
  {mn@eprint@\@tempb}{\@tempb:\@tempc}{\expandafter \expandafter \csname
  mn@eprint@\@tempb\endcsname \expandafter{\@tempc}}}

\bibitem[\protect\citeauthoryear{{Athanassoula}, {Romero-G{\'o}mez}, {Bosma}
  \& {Masdemont}}{{Athanassoula} et~al.}{2010}]{Athanassoula10}
{Athanassoula} E.,  {Romero-G{\'o}mez} M.,  {Bosma} A.,   {Masdemont} J.~J.,
  2010, \mn@doi [\mnras] {10.1111/j.1365-2966.2010.17010.x}, \href
  {https://ui.adsabs.harvard.edu/abs/2010MNRAS.407.1433A} {407, 1433}

\bibitem[\protect\citeauthoryear{{Bailer-Jones}}{{Bailer-Jones}}{2017}]{Bailer-Jones2017}
{Bailer-Jones} C. A.~L.,  2017, {Practical Bayesian Inference}

\bibitem[\protect\citeauthoryear{{Binney} \& {Tremaine}}{{Binney} \&
  {Tremaine}}{2008}]{Binney2008}
{Binney} J.,  {Tremaine} S.,  2008, {Galactic Dynamics: Second Edition}

\bibitem[\protect\citeauthoryear{{Bittner}, {Gadotti}, {Elmegreen},
  {Athanassoula}, {Elmegreen}, {Bosma}  \& {Mu{\~n}oz-Mateos}}{{Bittner}
  et~al.}{2017}]{Bittner2017}
{Bittner} A.,  {Gadotti} D.~A.,  {Elmegreen} B.~G.,  {Athanassoula} E.,
  {Elmegreen} D.~M.,  {Bosma} A.,   {Mu{\~n}oz-Mateos} J.-C.,  2017, \mn@doi
  [\mnras] {10.1093/mnras/stx1646}, \href
  {https://ui.adsabs.harvard.edu/abs/2017MNRAS.471.1070B} {471, 1070}

\bibitem[\protect\citeauthoryear{{Bizyaev}, {Kautsch}, {Mosenkov},
  {Reshetnikov}, {Sotnikova}, {Yablokova}  \& {Hillyer}}{{Bizyaev}
  et~al.}{2014}]{Bizyaev2014}
{Bizyaev} D.~V.,  {Kautsch} S.~J.,  {Mosenkov} A.~V.,  {Reshetnikov} V.~P.,
  {Sotnikova} N.~Y.,  {Yablokova} N.~V.,   {Hillyer} R.~W.,  2014, \mn@doi
  [\apj] {10.1088/0004-637X/787/1/24}, \href
  {https://ui.adsabs.harvard.edu/abs/2014ApJ...787...24B} {787, 24}

\bibitem[\protect\citeauthoryear{{Block}, {Bertin}, {Stockton}, {Grosbol},
  {Moorwood}  \& {Peletier}}{{Block} et~al.}{1994}]{1994A&A...288..365B}
{Block} D.~L.,  {Bertin} G.,  {Stockton} A.,  {Grosbol} P.,  {Moorwood}
  A.~F.~M.,   {Peletier} R.~F.,  1994, \aap, \href
  {https://ui.adsabs.harvard.edu/abs/1994A&A...288..365B} {288, 365}

\bibitem[\protect\citeauthoryear{{Buta} et~al.,}{{Buta}
  et~al.}{2015}]{Buta2015}
{Buta} R.~J.,  et~al., 2015, \mn@doi [\apjs] {10.1088/0067-0049/217/2/32},
  \href {https://ui.adsabs.harvard.edu/abs/2015ApJS..217...32B} {217, 32}

\bibitem[\protect\citeauthoryear{{Buzzo} et~al.,}{{Buzzo}
  et~al.}{2021}]{2021MNRAS.504.2146B}
{Buzzo} M.~L.,  et~al., 2021, \mn@doi [\mnras] {10.1093/mnras/stab941}, \href
  {https://ui.adsabs.harvard.edu/abs/2021MNRAS.504.2146B} {504, 2146}

\bibitem[\protect\citeauthoryear{{Cappellari} et~al.,}{{Cappellari}
  et~al.}{2013}]{Cappellari2013}
{Cappellari} M.,  et~al., 2013, \mn@doi [\mnras] {10.1093/mnras/stt644}, \href
  {https://ui.adsabs.harvard.edu/abs/2013MNRAS.432.1862C} {432, 1862}

\bibitem[\protect\citeauthoryear{{Casasola} et~al.,}{{Casasola}
  et~al.}{2017}]{Casasola2017}
{Casasola} V.,  et~al., 2017, \mn@doi [\aap] {10.1051/0004-6361/201731020},
  \href {https://ui.adsabs.harvard.edu/abs/2017A&A...605A..18C} {605, A18}

\bibitem[\protect\citeauthoryear{{Conselice}}{{Conselice}}{1997}]{Conselice1997}
{Conselice} C.~J.,  1997, \mn@doi [\pasp] {10.1086/134004}, \href
  {https://ui.adsabs.harvard.edu/abs/1997PASP..109.1251C} {109, 1251}

\bibitem[\protect\citeauthoryear{{Conselice}}{{Conselice}}{2006}]{Conselice2006}
{Conselice} C.~J.,  2006, \mn@doi [\mnras] {10.1111/j.1365-2966.2006.11114.x},
  \href {https://ui.adsabs.harvard.edu/abs/2006MNRAS.373.1389C} {373, 1389}

\bibitem[\protect\citeauthoryear{{D'Onofrio}, {Marziani}  \&
  {Chiosi}}{{D'Onofrio} et~al.}{2021}]{D'Onofrio2021}
{D'Onofrio} M.,  {Marziani} P.,   {Chiosi} C.,  2021, \mn@doi [Frontiers in
  Astronomy and Space Sciences] {10.3389/fspas.2021.694554}, \href
  {https://ui.adsabs.harvard.edu/abs/2021FrASS...8..157D} {8, 157}

\bibitem[\protect\citeauthoryear{{D'Souza}, {Kauffman}, {Wang}  \&
  {Vegetti}}{{D'Souza} et~al.}{2014}]{D'Souza2014}
{D'Souza} R.,  {Kauffman} G.,  {Wang} J.,   {Vegetti} S.,  2014, \mn@doi
  [\mnras] {10.1093/mnras/stu1194}, \href
  {https://ui.adsabs.harvard.edu/abs/2014MNRAS.443.1433D} {443, 1433}

\bibitem[\protect\citeauthoryear{{Davis}, {Berrier}, {Shields}, {Kennefick},
  {Kennefick}, {Seigar}, {Lacy}  \& {Puerari}}{{Davis}
  et~al.}{2012}]{Davis2012}
{Davis} B.~L.,  {Berrier} J.~C.,  {Shields} D.~W.,  {Kennefick} J.,
  {Kennefick} D.,  {Seigar} M.~S.,  {Lacy} C. H.~S.,   {Puerari} I.,  2012,
  \mn@doi [\apjs] {10.1088/0067-0049/199/2/33}, \href
  {https://ui.adsabs.harvard.edu/abs/2012ApJS..199...33D} {199, 33}

\bibitem[\protect\citeauthoryear{{Davis} et~al.,}{{Davis}
  et~al.}{2015}]{Davis2015}
{Davis} B.~L.,  et~al., 2015, \mn@doi [\apjl] {10.1088/2041-8205/802/1/L13},
  \href {https://ui.adsabs.harvard.edu/abs/2015ApJ...802L..13D} {802, L13}

\bibitem[\protect\citeauthoryear{{D{\'\i}az-Garc{\'\i}a}, {Salo}, {Knapen}  \&
  {Herrera-Endoqui}}{{D{\'\i}az-Garc{\'\i}a} et~al.}{2019}]{Diaz-Garcia2019}
{D{\'\i}az-Garc{\'\i}a} S.,  {Salo} H.,  {Knapen} J.~H.,   {Herrera-Endoqui}
  M.,  2019, \mn@doi [\aap] {10.1051/0004-6361/201936000}, \href
  {https://ui.adsabs.harvard.edu/abs/2019A&A...631A..94D} {631, A94}

\bibitem[\protect\citeauthoryear{{Elmegreen}}{{Elmegreen}}{1981}]{1981ApJS...47..229E}
{Elmegreen} D.~M.,  1981, \mn@doi [\apjs] {10.1086/190757}, \href
  {https://ui.adsabs.harvard.edu/abs/1981ApJS...47..229E} {47, 229}

\bibitem[\protect\citeauthoryear{{Elmegreen}}{{Elmegreen}}{1990}]{Elmegreen1990}
{Elmegreen} B.~G.,  1990, \mn@doi [Annals of the New York Academy of Sciences]
  {10.1111/j.1749-6632.1990.tb27410.x}, \href
  {https://ui.adsabs.harvard.edu/abs/1990NYASA.596...40E} {596, 40}

\bibitem[\protect\citeauthoryear{{Elmegreen} \& {Elmegreen}}{{Elmegreen} \&
  {Elmegreen}}{1982}]{Elmegreen1982}
{Elmegreen} D.~M.,  {Elmegreen} B.~G.,  1982, \mn@doi [\mnras]
  {10.1093/mnras/201.4.1021}, \href
  {https://ui.adsabs.harvard.edu/abs/1982MNRAS.201.1021E} {201, 1021}

\bibitem[\protect\citeauthoryear{{Elmegreen} \& {Elmegreen}}{{Elmegreen} \&
  {Elmegreen}}{1984}]{1984ApJS...54..127E}
{Elmegreen} D.~M.,  {Elmegreen} B.~G.,  1984, \mn@doi [\apjs] {10.1086/190922},
  \href {https://ui.adsabs.harvard.edu/abs/1984ApJS...54..127E} {54, 127}

\bibitem[\protect\citeauthoryear{{Elmegreen} \& {Elmegreen}}{{Elmegreen} \&
  {Elmegreen}}{1987}]{Elmegreen1987}
{Elmegreen} D.~M.,  {Elmegreen} B.~G.,  1987, \mn@doi [\apj] {10.1086/165034},
  \href {https://ui.adsabs.harvard.edu/abs/1987ApJ...314....3E} {314, 3}

\bibitem[\protect\citeauthoryear{{Erwin}}{{Erwin}}{2015}]{Erwin2015}
{Erwin} P.,  2015, \mn@doi [\apj] {10.1088/0004-637X/799/2/226}, \href
  {https://ui.adsabs.harvard.edu/abs/2015ApJ...799..226E} {799, 226}

\bibitem[\protect\citeauthoryear{{Erwin} et~al.,}{{Erwin}
  et~al.}{2021}]{Erwin2021}
{Erwin} P.,  et~al., 2021, \mn@doi [\mnras] {10.1093/mnras/stab126}, \href
  {https://ui.adsabs.harvard.edu/abs/2021MNRAS.502.2446E} {502, 2446}

\bibitem[\protect\citeauthoryear{{Fisher} \& {Drory}}{{Fisher} \&
  {Drory}}{2010}]{Fisher2010}
{Fisher} D.~B.,  {Drory} N.,  2010, \mn@doi [\apj]
  {10.1088/0004-637X/716/2/942}, \href
  {https://ui.adsabs.harvard.edu/abs/2010ApJ...716..942F} {716, 942}

\bibitem[\protect\citeauthoryear{{Font}, {Beckman}, {James}  \&
  {Patsis}}{{Font} et~al.}{2019}]{Font2019}
{Font} J.,  {Beckman} J.~E.,  {James} P.~A.,   {Patsis} P.~A.,  2019, \mn@doi
  [\mnras] {10.1093/mnras/sty2983}, \href
  {https://ui.adsabs.harvard.edu/abs/2019MNRAS.482.5362F} {482, 5362}

\bibitem[\protect\citeauthoryear{{Forgan}, {Ram{\'o}n-Fox}  \&
  {Bonnell}}{{Forgan} et~al.}{2018}]{Forgan2018}
{Forgan} D.~H.,  {Ram{\'o}n-Fox} F.~G.,   {Bonnell} I.~A.,  2018, \mn@doi
  [\mnras] {10.1093/mnras/sty331}, \href
  {https://ui.adsabs.harvard.edu/abs/2018MNRAS.476.2384F} {476, 2384}

\bibitem[\protect\citeauthoryear{{Freeman}}{{Freeman}}{1970}]{Freeman1970}
{Freeman} K.~C.,  1970, \mn@doi [\apj] {10.1086/150474}, \href
  {https://ui.adsabs.harvard.edu/abs/1970ApJ...160..811F} {160, 811}

\bibitem[\protect\citeauthoryear{{Gadotti}}{{Gadotti}}{2008}]{Gadotti2008}
{Gadotti} D.~A.,  2008, \mn@doi [\mnras] {10.1111/j.1365-2966.2007.12723.x},
  \href {https://ui.adsabs.harvard.edu/abs/2008MNRAS.384..420G} {384, 420}

\bibitem[\protect\citeauthoryear{{Gadotti}}{{Gadotti}}{2009}]{Gadotti2009}
{Gadotti} D.~A.,  2009, \mn@doi [\mnras] {10.1111/j.1365-2966.2008.14257.x},
  \href {https://ui.adsabs.harvard.edu/abs/2009MNRAS.393.1531G} {393, 1531}

\bibitem[\protect\citeauthoryear{{Gadotti} et~al.,}{{Gadotti}
  et~al.}{2020}]{Gadotti2020}
{Gadotti} D.~A.,  et~al., 2020, \mn@doi [\aap] {10.1051/0004-6361/202038448},
  \href {https://ui.adsabs.harvard.edu/abs/2020A&A...643A..14G} {643, A14}

\bibitem[\protect\citeauthoryear{{Gao} \& {Ho}}{{Gao} \& {Ho}}{2017}]{Gao2017}
{Gao} H.,  {Ho} L.~C.,  2017, \mn@doi [\apj] {10.3847/1538-4357/aa7da4}, \href
  {https://ui.adsabs.harvard.edu/abs/2017ApJ...845..114G} {845, 114}

\bibitem[\protect\citeauthoryear{{Gong}, {Mao}, {Gao}  \& {Yu}}{{Gong}
  et~al.}{2023}]{2023ApJS..267...26G}
{Gong} J.-Y.,  {Mao} Y.-W.,  {Gao} H.,   {Yu} S.-Y.,  2023, \mn@doi [\apjs]
  {10.3847/1538-4365/acd554}, \href
  {https://ui.adsabs.harvard.edu/abs/2023ApJS..267...26G} {267, 26}

\bibitem[\protect\citeauthoryear{{Hamilton}}{{Hamilton}}{2023}]{Hamilton2023}
{Hamilton} C.,  2023, \mn@doi [arXiv e-prints] {10.48550/arXiv.2302.06602},
  \href {https://ui.adsabs.harvard.edu/abs/2023arXiv230206602H} {p.
  arXiv:2302.06602}

\bibitem[\protect\citeauthoryear{{Hart}, {Bamford}, {Casteels}, {Kruk},
  {Lintott}  \& {Masters}}{{Hart} et~al.}{2017}]{Hart2017}
{Hart} R.~E.,  {Bamford} S.~P.,  {Casteels} K. R.~V.,  {Kruk} S.~J.,  {Lintott}
  C.~J.,   {Masters} K.~L.,  2017, \mn@doi [\mnras] {10.1093/mnras/stx581},
  \href {https://ui.adsabs.harvard.edu/abs/2017MNRAS.468.1850H} {468, 1850}

\bibitem[\protect\citeauthoryear{{Honig} \& {Reid}}{{Honig} \&
  {Reid}}{2015}]{Honig2015}
{Honig} Z.~N.,  {Reid} M.~J.,  2015, \mn@doi [\apj]
  {10.1088/0004-637X/800/1/53}, \href
  {https://ui.adsabs.harvard.edu/abs/2015ApJ...800...53H} {800, 53}

\bibitem[\protect\citeauthoryear{{Hunter} \& {Elmegreen}}{{Hunter} \&
  {Elmegreen}}{2006}]{Hunter2006}
{Hunter} D.~A.,  {Elmegreen} B.~G.,  2006, \mn@doi [\apjs] {10.1086/498096},
  \href {https://ui.adsabs.harvard.edu/abs/2006ApJS..162...49H} {162, 49}

\bibitem[\protect\citeauthoryear{{Joye} \& {Mandel}}{{Joye} \&
  {Mandel}}{2003}]{Joye2003}
{Joye} W.~A.,  {Mandel} E.,  2003, in {Payne} H.~E.,  {Jedrzejewski} R.~I.,
  {Hook} R.~N.,  eds,  Astronomical Society of the Pacific Conference Series
  Vol. 295, Astronomical Data Analysis Software and Systems XII. p.~489

\bibitem[\protect\citeauthoryear{{Kendall}, {Kennicutt}  \& {Clarke}}{{Kendall}
  et~al.}{2011}]{Kendall2011}
{Kendall} S.,  {Kennicutt} R.~C.,   {Clarke} C.,  2011, \mn@doi [\mnras]
  {10.1111/j.1365-2966.2011.18422.x}, \href
  {https://ui.adsabs.harvard.edu/abs/2011MNRAS.414..538K} {414, 538}

\bibitem[\protect\citeauthoryear{{Kendall}, {Clarke}  \& {Kennicutt}}{{Kendall}
  et~al.}{2015}]{Kendall2015}
{Kendall} S.,  {Clarke} C.,   {Kennicutt} R.~C.,  2015, \mn@doi [\mnras]
  {10.1093/mnras/stu2431}, \href
  {https://ui.adsabs.harvard.edu/abs/2015MNRAS.446.4155K} {446, 4155}

\bibitem[\protect\citeauthoryear{{Kennicutt}}{{Kennicutt}}{1981}]{Kennicutt1981}
{Kennicutt} R.~C. J.,  1981, \mn@doi [\aj] {10.1086/113064}, \href
  {https://ui.adsabs.harvard.edu/abs/1981AJ.....86.1847K} {86, 1847}

\bibitem[\protect\citeauthoryear{{Kregel} \& {van der Kruit}}{{Kregel} \& {van
  der Kruit}}{2004}]{Kregel2004}
{Kregel} M.,  {van der Kruit} P.~C.,  2004, \mn@doi [\mnras]
  {10.1111/j.1365-2966.2004.08307.x}, \href
  {https://ui.adsabs.harvard.edu/abs/2004MNRAS.355..143K} {355, 143}

\bibitem[\protect\citeauthoryear{{Laine} et~al.,}{{Laine}
  et~al.}{2014}]{Laine2014}
{Laine} J.,  et~al., 2014, \mn@doi [\mnras] {10.1093/mnras/stu628}, \href
  {https://ui.adsabs.harvard.edu/abs/2014MNRAS.441.1992L} {441, 1992}

\bibitem[\protect\citeauthoryear{{Lang}, {Hogg}  \& {Mykytyn}}{{Lang}
  et~al.}{2016}]{Lang2016}
{Lang} D.,  {Hogg} D.~W.,   {Mykytyn} D.,  2016, {The Tractor: Probabilistic
  astronomical source detection and measurement}, Astrophysics Source Code
  Library, record ascl:1604.008 (\mn@eprint {ascl} {1604.008})

\bibitem[\protect\citeauthoryear{{L{\"a}sker}, {Ferrarese}  \& {van de
  Ven}}{{L{\"a}sker} et~al.}{2014}]{Lasker2014}
{L{\"a}sker} R.,  {Ferrarese} L.,   {van de Ven} G.,  2014, \mn@doi [\apj]
  {10.1088/0004-637X/780/1/69}, \href
  {https://ui.adsabs.harvard.edu/abs/2014ApJ...780...69L} {780, 69}

\bibitem[\protect\citeauthoryear{{Lingard} et~al.,}{{Lingard}
  et~al.}{2020}]{Lingard2020}
{Lingard} T.~K.,  et~al., 2020, \mn@doi [\apj] {10.3847/1538-4357/ab9d83},
  \href {https://ui.adsabs.harvard.edu/abs/2020ApJ...900..178L} {900, 178}

\bibitem[\protect\citeauthoryear{{Lingard} et~al.,}{{Lingard}
  et~al.}{2021}]{Lingard2021}
{Lingard} T.,  et~al., 2021, \mn@doi [\mnras] {10.1093/mnras/stab1072}, \href
  {https://ui.adsabs.harvard.edu/abs/2021MNRAS.504.3364L} {504, 3364}

\bibitem[\protect\citeauthoryear{{Makarov}, {Prugniel}, {Terekhova}, {Courtois}
   \& {Vauglin}}{{Makarov} et~al.}{2014}]{Makarov2014}
{Makarov} D.,  {Prugniel} P.,  {Terekhova} N.,  {Courtois} H.,   {Vauglin} I.,
  2014, \mn@doi [\aap] {10.1051/0004-6361/201423496}, \href
  {https://ui.adsabs.harvard.edu/abs/2014A&A...570A..13M} {570, A13}

\bibitem[\protect\citeauthoryear{{Marchuk} et~al.,}{{Marchuk}
  et~al.}{2022}]{Marchuk2022}
{Marchuk} A.~A.,  et~al., 2022, \mn@doi [\mnras] {10.1093/mnras/stac599}, \href
  {https://ui.adsabs.harvard.edu/abs/2022MNRAS.512.1371M} {512, 1371}

\bibitem[\protect\citeauthoryear{{Marchuk}, {Mosenkov}, {Chugunov}, {Kostiuk},
  {Skryabina}  \& {Reshetnikov}}{{Marchuk} et~al.}{2023}]{Marchuk2023}
{Marchuk} A.~A.,  {Mosenkov} A.~V.,  {Chugunov} I.~V.,  {Kostiuk} V.~S.,
  {Skryabina} M.~N.,   {Reshetnikov} V.~P.,  2023, arXiv e-prints, \href
  {https://ui.adsabs.harvard.edu/abs/2023arXiv230916232M} {p. arXiv:2309.16232}

\bibitem[\protect\citeauthoryear{{Mart{\'\i}nez-Garc{\'\i}a},
  {Gonz{\'a}lez-L{\'o}pezlira}  \& {Puerari}}{{Mart{\'\i}nez-Garc{\'\i}a}
  et~al.}{2023}]{2023MNRAS.524...18M}
{Mart{\'\i}nez-Garc{\'\i}a} E.~E.,  {Gonz{\'a}lez-L{\'o}pezlira} R.~A.,
  {Puerari} I.,  2023, \mn@doi [\mnras] {10.1093/mnras/stad1805}, \href
  {https://ui.adsabs.harvard.edu/abs/2023MNRAS.524...18M} {524, 18}

\bibitem[\protect\citeauthoryear{{Masters} \& {Galaxy Zoo Team}}{{Masters} \&
  {Galaxy Zoo Team}}{2020}]{Masters2020}
{Masters} K.~L.,  {Galaxy Zoo Team} 2020, \mn@doi [Proceedings of the
  International Astronomical Union] {10.1017/S1743921319008615}, \href
  {https://ui.adsabs.harvard.edu/abs/2020IAUS..353..205M} {353, 205}

\bibitem[\protect\citeauthoryear{{M{\'e}ndez-Abreu} et~al.,}{{M{\'e}ndez-Abreu}
  et~al.}{2017}]{Mendez-Abreu2017}
{M{\'e}ndez-Abreu} J.,  et~al., 2017, \mn@doi [\aap]
  {10.1051/0004-6361/201629525}, \href
  {https://ui.adsabs.harvard.edu/abs/2017A&A...598A..32M} {598, A32}

\bibitem[\protect\citeauthoryear{{Minchev}, {Famaey}, {Quillen}, {Di Matteo},
  {Combes}, {Vlaji{\'c}}, {Erwin}  \& {Bland-Hawthorn}}{{Minchev}
  et~al.}{2012}]{Minchev12}
{Minchev} I.,  {Famaey} B.,  {Quillen} A.~C.,  {Di Matteo} P.,  {Combes} F.,
  {Vlaji{\'c}} M.,  {Erwin} P.,   {Bland-Hawthorn} J.,  2012, \mn@doi [\aap]
  {10.1051/0004-6361/201219198}, \href
  {https://ui.adsabs.harvard.edu/abs/2012A&A...548A.126M} {548, A126}

\bibitem[\protect\citeauthoryear{{Mor{\'e}}}{{Mor{\'e}}}{1978}]{More1978}
{Mor{\'e}} J.~J.,  1978, in Watson G.~A.,  ed., Numerical Analysis. Springer
  Berlin Heidelberg, Berlin, Heidelberg, pp 105--116

\bibitem[\protect\citeauthoryear{{Mosenkov}, {Savchenko}, {Smirnov}  \&
  {Camps}}{{Mosenkov} et~al.}{2021}]{Mosenkov2021}
{Mosenkov} A.~V.,  {Savchenko} S.~S.,  {Smirnov} A.~A.,   {Camps} P.,  2021,
  \mn@doi [\mnras] {10.1093/mnras/stab2445}, \href
  {https://ui.adsabs.harvard.edu/abs/2021MNRAS.507.5246M} {507, 5246}

\bibitem[\protect\citeauthoryear{{O'Neil} \& {Bothun}}{{O'Neil} \&
  {Bothun}}{2000}]{O'Neil2000}
{O'Neil} K.,  {Bothun} G.,  2000, \mn@doi [\apj] {10.1086/308322}, \href
  {https://ui.adsabs.harvard.edu/abs/2000ApJ...529..811O} {529, 811}

\bibitem[\protect\citeauthoryear{{Papaderos}, {Breda}, {Humphrey}, {Michel
  Gomes}, {Ziegler}  \& {Pappalardo}}{{Papaderos} et~al.}{2022}]{Papaderos2022}
{Papaderos} P.,  {Breda} I.,  {Humphrey} A.,  {Michel Gomes} J.,  {Ziegler}
  B.~L.,   {Pappalardo} C.,  2022, \mn@doi [\aap]
  {10.1051/0004-6361/202140641}, \href
  {https://ui.adsabs.harvard.edu/abs/2022A&A...658A..74P} {658, A74}

\bibitem[\protect\citeauthoryear{{Peng}, {Ho}, {Impey}  \& {Rix}}{{Peng}
  et~al.}{2002}]{Peng2002}
{Peng} C.~Y.,  {Ho} L.~C.,  {Impey} C.~D.,   {Rix} H.-W.,  2002, \mn@doi [\aj]
  {10.1086/340952}, \href
  {https://ui.adsabs.harvard.edu/abs/2002AJ....124..266P} {124, 266}

\bibitem[\protect\citeauthoryear{{Peng}, {Ho}, {Impey}  \& {Rix}}{{Peng}
  et~al.}{2010}]{Peng2010}
{Peng} C.~Y.,  {Ho} L.~C.,  {Impey} C.~D.,   {Rix} H.-W.,  2010, \mn@doi [\aj]
  {10.1088/0004-6256/139/6/2097}, \href
  {https://ui.adsabs.harvard.edu/abs/2010AJ....139.2097P} {139, 2097}

\bibitem[\protect\citeauthoryear{{P{\'e}rez-Villegas}, {Pichardo}  \&
  {Moreno}}{{P{\'e}rez-Villegas} et~al.}{2015}]{Perez-Villegas2015}
{P{\'e}rez-Villegas} A.,  {Pichardo} B.,   {Moreno} E.,  2015, \mn@doi [\apj]
  {10.1088/0004-637X/809/2/170}, \href
  {https://ui.adsabs.harvard.edu/abs/2015ApJ...809..170P} {809, 170}

\bibitem[\protect\citeauthoryear{{Pohlen} \& {Trujillo}}{{Pohlen} \&
  {Trujillo}}{2006}]{Pohlen2006}
{Pohlen} M.,  {Trujillo} I.,  2006, \mn@doi [\aap]
  {10.1051/0004-6361:20064883}, \href
  {https://ui.adsabs.harvard.edu/abs/2006A&A...454..759P} {454, 759}

\bibitem[\protect\citeauthoryear{Roberts \& Haynes}{Roberts \&
  Haynes}{1994}]{Roberts1994}
Roberts M.~S.,  Haynes M.~P.,  1994, \mn@doi [\araa]
  {10.1146/annurev.aa.32.090194.000555}, 32, 115

\bibitem[\protect\citeauthoryear{{Salo} et~al.,}{{Salo}
  et~al.}{2015}]{Salo2015}
{Salo} H.,  et~al., 2015, \mn@doi [\apjs] {10.1088/0067-0049/219/1/4}, \href
  {https://ui.adsabs.harvard.edu/abs/2015ApJS..219....4S} {219, 4}

\bibitem[\protect\citeauthoryear{{Savchenko} \& {Reshetnikov}}{{Savchenko} \&
  {Reshetnikov}}{2013}]{Savchenko2013}
{Savchenko} S.~S.,  {Reshetnikov} V.~P.,  2013, \mn@doi [\mnras]
  {10.1093/mnras/stt1627}, \href
  {https://ui.adsabs.harvard.edu/abs/2013MNRAS.436.1074S} {436, 1074}

\bibitem[\protect\citeauthoryear{{Savchenko}, {Marchuk}, {Mosenkov}  \&
  {Grishunin}}{{Savchenko} et~al.}{2020}]{Savchenko2020}
{Savchenko} S.,  {Marchuk} A.,  {Mosenkov} A.,   {Grishunin} K.,  2020, \mn@doi
  [\mnras] {10.1093/mnras/staa258}, \href
  {https://ui.adsabs.harvard.edu/abs/2020MNRAS.493..390S} {493, 390}

\bibitem[\protect\citeauthoryear{{Sersic}}{{Sersic}}{1968}]{Sersic1968}
{Sersic} J.~L.,  1968, {Atlas de Galaxias Australes}

\bibitem[\protect\citeauthoryear{{Sheth} et~al.,}{{Sheth}
  et~al.}{2010}]{Sheth2010}
{Sheth} K.,  et~al., 2010, \mn@doi [\pasp] {10.1086/657638}, \href
  {https://ui.adsabs.harvard.edu/abs/2010PASP..122.1397S} {122, 1397}

\bibitem[\protect\citeauthoryear{{Simard}, {Mendel}, {Patton}, {Ellison}  \&
  {McConnachie}}{{Simard} et~al.}{2011}]{Simard2011}
{Simard} L.,  {Mendel} J.~T.,  {Patton} D.~R.,  {Ellison} S.~L.,
  {McConnachie} A.~W.,  2011, \mn@doi [\apjs] {10.1088/0067-0049/196/1/11},
  \href {https://ui.adsabs.harvard.edu/abs/2011ApJS..196...11S} {196, 11}

\bibitem[\protect\citeauthoryear{{Smirnov} \& {Savchenko}}{{Smirnov} \&
  {Savchenko}}{2020}]{Smirnov2020}
{Smirnov} A.~A.,  {Savchenko} S.~S.,  2020, \mn@doi [\mnras]
  {10.1093/mnras/staa2892}, \href
  {https://ui.adsabs.harvard.edu/abs/2020MNRAS.499..462S} {499, 462}

\bibitem[\protect\citeauthoryear{{Sonnenfeld}}{{Sonnenfeld}}{2022}]{Sonnenfeld2022}
{Sonnenfeld} A.,  2022, \mn@doi [\aap] {10.1051/0004-6361/202142786}, \href
  {https://ui.adsabs.harvard.edu/abs/2022A&A...659A.141S} {659, A141}

\bibitem[\protect\citeauthoryear{{Vika}, {Driver}, {Cameron}, {Kelvin}  \&
  {Robotham}}{{Vika} et~al.}{2012}]{Vika2012}
{Vika} M.,  {Driver} S.~P.,  {Cameron} E.,  {Kelvin} L.,   {Robotham} A.,
  2012, \mn@doi [\mnras] {10.1111/j.1365-2966.2011.19881.x}, \href
  {https://ui.adsabs.harvard.edu/abs/2012MNRAS.419.2264V} {419, 2264}

\bibitem[\protect\citeauthoryear{{Yu} \& {Ho}}{{Yu} \& {Ho}}{2019}]{Yu2019}
{Yu} S.-Y.,  {Ho} L.~C.,  2019, \mn@doi [\apj] {10.3847/1538-4357/aaf895},
  \href {https://ui.adsabs.harvard.edu/abs/2019ApJ...871..194Y} {871, 194}

\bibitem[\protect\citeauthoryear{{de Vaucouleurs}}{{de
  Vaucouleurs}}{1948}]{deVaucouleurs1948}
{de Vaucouleurs} G.,  1948, Annales d'Astrophysique, \href
  {https://ui.adsabs.harvard.edu/abs/1948AnAp...11..247D} {11, 247}

\makeatother
\end{thebibliography}

\section*{Appendix A}
\label{sec:appendix}
\begin{figure*}
\centering
\begin{minipage}[c]{0.49\textwidth}
\centering
\includegraphics[width=\textwidth]{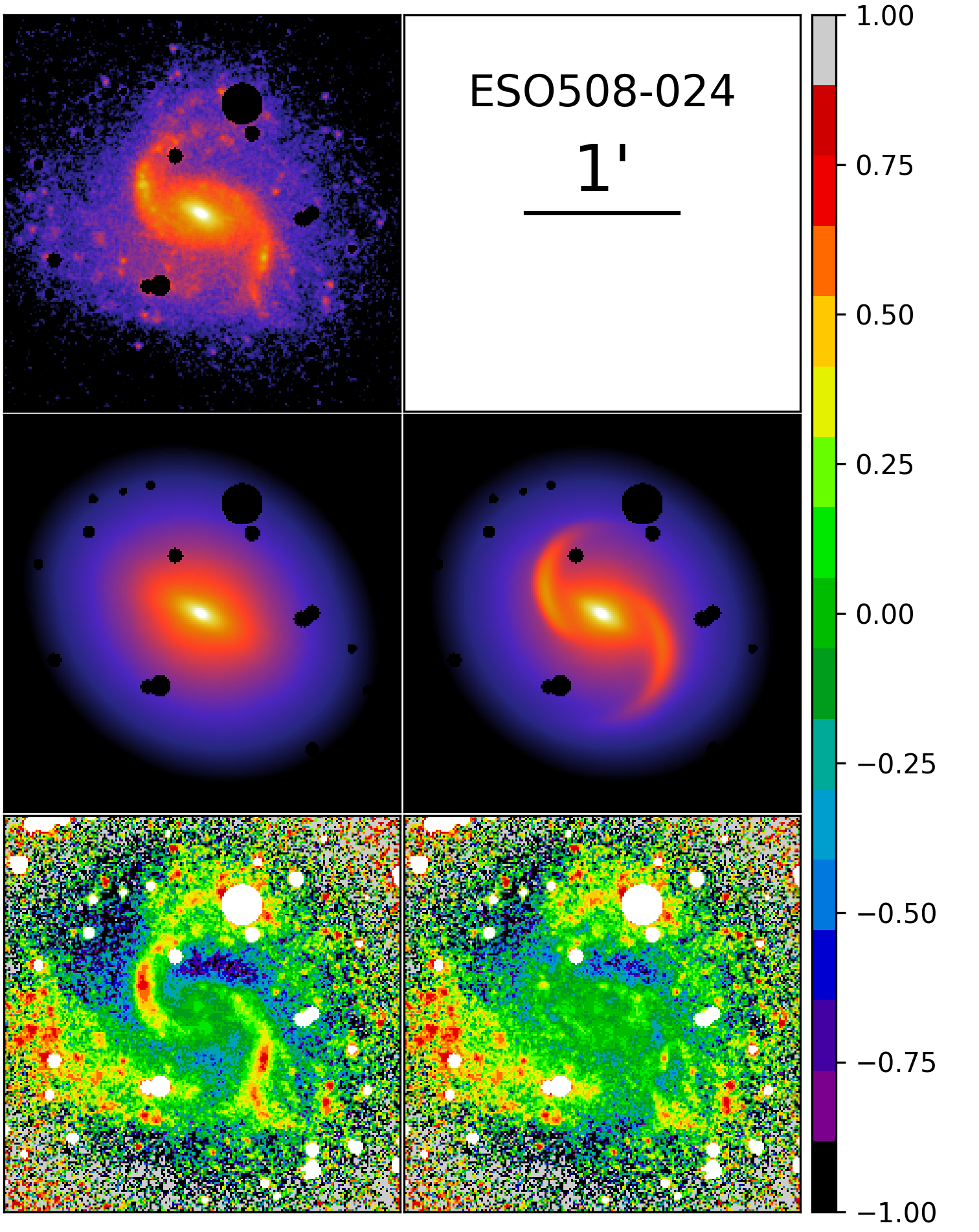}
\end{minipage}
\hfill
\begin{minipage}[c]{0.49\textwidth}
\centering
\includegraphics[width=\textwidth]{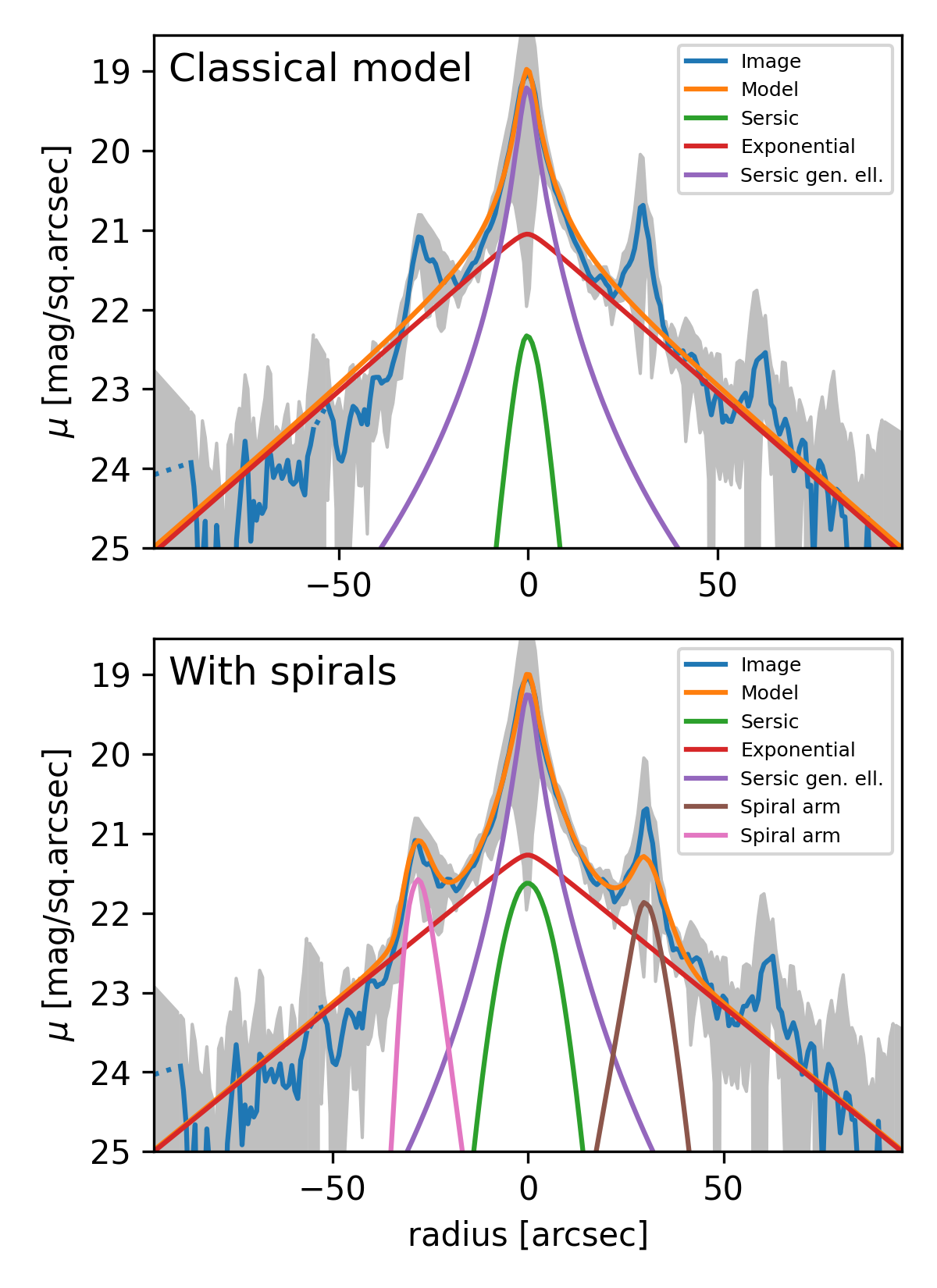}
\end{minipage}

\medskip

\begin{minipage}[t]{0.49\textwidth}
\centering
\vspace{0pt}
Components besides spiral arms\\
\begin{tabular}{lccc}
\hline
Component & Parameter & Units & Value \\
\hline
Disc& $I_0$ & mag/arcsec$^2$ & $21.18 \pm 0.01$ \\
(Exponential)& $h$ & arcsec & $27.15 \pm 0.10$ \\
& Frac. & & $0.81$ \\
\hline
Bulge& $I_\text{e}$ & mag/arcsec$^2$ & $22.30 \pm 0.05$ \\
(Sersic)& $R_\text{e}$ & arcsec & $7.50 \pm 0.00$ \\
& $n$ & & $0.52 \pm 0.06$ \\
& Frac. & & $0.03$ \\
\hline
Bar& $I_\text{e}$ & mag/arcsec$^2$ & $21.00 \pm 0.03$ \\
(Sersic gen. ell.)& $R_\text{e}$ & arcsec & $7.50 \pm 0.00$ \\
& $n$ & & $3.05 \pm 0.08$ \\
& Frac. & & $0.08$ \\
\hline
\end{tabular}
\end{minipage}
\hfill
\begin{minipage}[t]{0.49\textwidth}
\centering
\vspace{0pt}
\includegraphics[width=\textwidth]{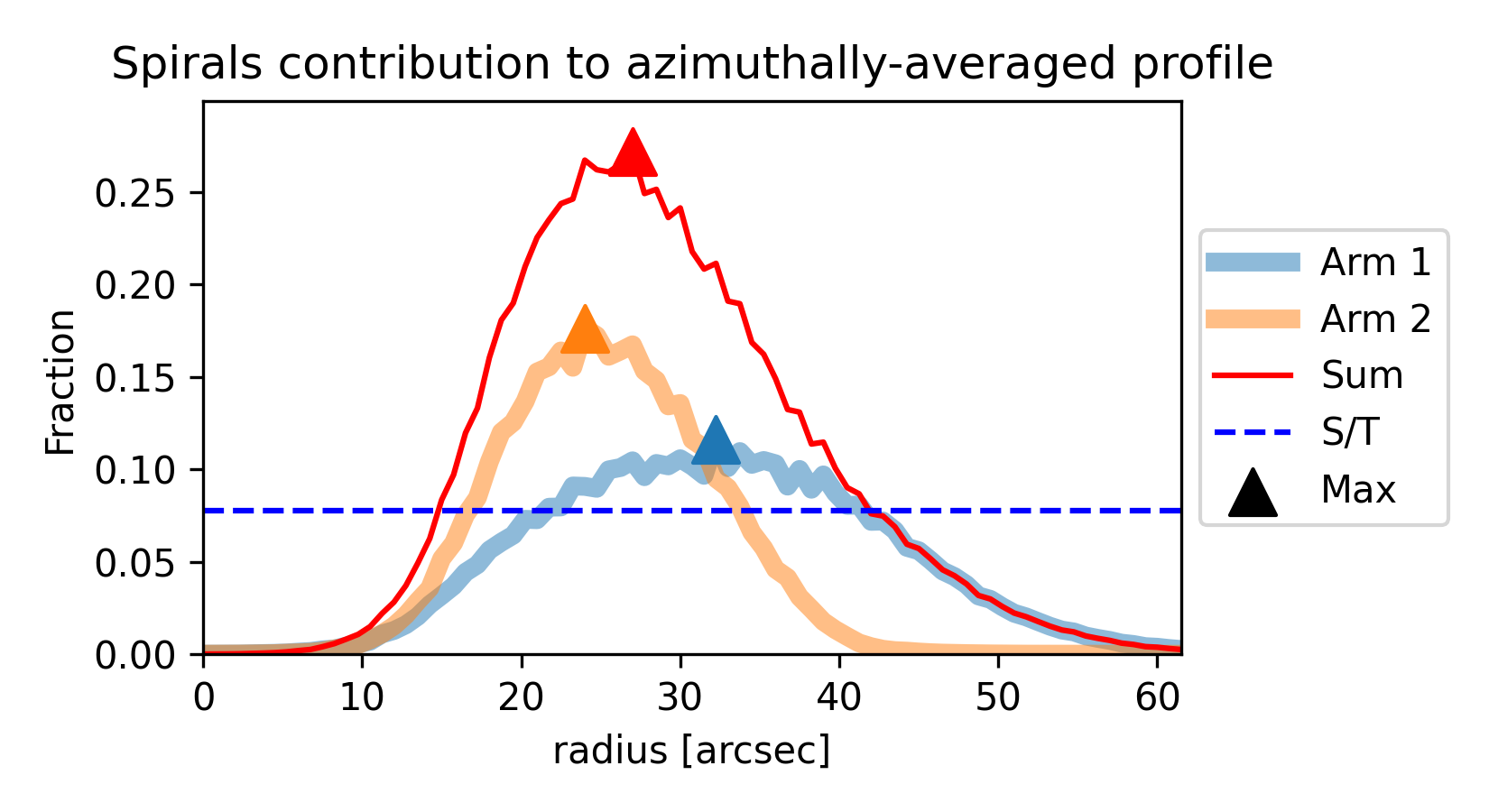}
Spiral arms parameters\\
\begin{tabular}{lcc}
\hline
& Arm 1 & Arm 2 \\
\hline
$I_\text{max}$,~mag/arcsec$^2$& $21.51 \pm 0.02$ & $21.04 \pm 0.02$ \\
$h_s$,~arcsec& $3.21 \pm 0.08$ & $1.71 \pm 0.03$ \\
Width,~arcsec& $6.62 \pm 0.34$ & $6.35 \pm 0.16$ \\
$\upmu$,~deg& $20.0 \pm 1.1$ & $16.8 \pm 1.0$ \\
$\Delta \upmu$,~deg& $6.3 \pm 1.1$ & $4.9 \pm 1.0$ \\
Frac & 0.04 & 0.04 \\
\hline
\end{tabular}
\end{minipage}
\caption{Decomposition results for ESO508-024. \textbf{Images:} Original image, models, and residuals (similar to Fig.~\ref{fig:NGC5247}) are at the top left. Surface brightness profiles of image, model, and individual components along the major axis are at the top right. A plot of the spiral arms contribution to the azimuthally-averaged profile of galaxy (similar to Fig.~\ref{fig:az_examples} but less detailed) is at the lower right. \textbf{Tables:} List of ``classical'' components and their parameters in the model with spiral arms is on the left. A list of the spiral arms with their parameters is on the right.}
\label{fig:appendix_example}
\end{figure*}

In Fig.~\ref{fig:appendix_example}, we present the summarised decomposition results for ESO 508-024 galaxy, showing 2D images of galaxy and models, major-axis surface brightness profiles, spiral arms contribution to azimuthally-averaged profile, as well as tables with the list of parameters of components. Similar figures for all 29 galaxies are available in the online material.

\centering

\label{lastpage}

\end{document}